\documentclass{aa}  

\usepackage{epstopdf}
\usepackage{multirow}
\usepackage{textcomp}
\usepackage{graphicx}
\usepackage{amsmath}
\usepackage{txfonts}
\begin{document}

   \title{New giant radio sources and underluminous radio halos in two galaxy clusters}

\authorrunning{V. Cuciti et al.}
   \author{V. Cuciti
          \inst{1,2*},
          G. Brunetti\inst{2}, R. van Weeren\inst{3}, A. Bonafede\inst{2}, D. Dallacasa\inst{1,2}, R. Cassano\inst{2}, T. Venturi\inst{2}, R. Kale\inst{4}
          }

   \institute{Dipartimento di Fisica e Astronomia, Universit\`{a} di Bologna, via P. Gobetti 93/2, 40129 Bologna, Italy
         \and
                      INAF-Istituto di Radioastronomia, via P. Gobetti 101, 40129 Bologna, Italy
             \and
             Harvard-Smithsonian Center for Astrophysics, 60 Garden Street, Cambridge, MA 02138, USA
             \and
             National Centre for Radio Astrophysics, Tata Institute of Fundamental Research, Post Bag 3, Pune 411007, India\\             
             *\email vcuciti@ira.inaf.it}

   \date{Received September 15, 1996; accepted March 16, 1997}

 
  \abstract
   {}
   {The aim of this work is to analyse the radio properties of the massive and dynamical disturbed clusters Abell 1451 and Zwcl 0634.1+4750, especially focusing on the possible presence of diffuse emission.}
   {We present new GMRT 320 MHz and JVLA 1.5 GHz observations of these two clusters. 
   }
   {We found that both Abell 1451 and Zwcl 0634.1+4750 host a radio halo with a typical spectrum ($\alpha\sim1-1.3$). Similarly to a few other cases reported in the recent literature, these radio halos are significantly fainter in radio luminosity with respect to the current radio power-mass correlations and they are smaller than classical giant radio halos. These underluminous sources might contribute to shed light on the complex mechanisms of formation and evolution of radio halos. 
Furthermore, we detected a candidate radio relic at large distance from the cluster center in Abell 1451 and a peculiar head tail radio galaxy in Zwcl 0634.1+4750, which might be interacting with a shock front.}
   {}

   \keywords{Galaxies: clusters: individual --
                Galaxies: clusters: intracluster medium --
                Radiation mechanisms: non-thermal
               }

   \maketitle
%

\section{Introduction}
\label{Sec:intro}
Distorted radio galaxies with complex morphologies are often observed in galaxy clusters \citep[e.g.][]{willson70,hill71,wellington73,eilek84,o'dea&owen86,pinkey93,feretti01,ferettiventuri02,lane02,giacintucci09,mao09,owen14}. These spectacular sources are the result of the interaction between the radio galaxies and the dense surrounding ICM, which causes the bending of the jets and lobes of the radio galaxies \citep{jaffe73,begelman79,owen76,miley80}. 
In addition, some galaxy clusters host diffuse synchrotron emission associated with relativistic particles and magnetic fields in the intra cluster medium (ICM). Depending on their location and morphology, these cluster-scale radio sources are referred to as radio halos or relics. radio halos are located at the center of clusters and their morphology roughly follows the X-ray emission from the thermal ICM. Radio relics are situated at the periphery of the clusters and they typically have elongated or arc-like shapes \citep{feretti12}. 
There is compelling evidence that radio halos and relics are hosted by dynamically disturbed systems, while clusters with no cluster-scale radio emission are generally relaxed \citep{cassano10,cassano12,cassano13,cuciti15,degasperin14}. These sources suggest that part of the gravitational energy injected in form of shocks and large scale gas motions is drained into non-thermal components trough a hierarchy of complex mechanisms \citep{brunettijones14}.

In the current theoretical picture, radio halos form via the acceleration of relativistic electrons by means of cluster-wide turbulence injected in the ICM during merging events \citep{brunetti01,petrosian01}. Alternatively, it has been proposed that radio halos are caused by secondary electrons generated by proton-proton collisions \citep{dennison80,blasicolafrancesco99}, however the non-detection of galaxy clusters in the gamma-rays disfavours this possibility \citep{ackermann10,ackermann14,ackermann16,jeltema11, brunetti12, zandanel14,brunetti17}. Models based on the combination of the two mechanisms where secondary electrons are reaccelerated by turbulence have also been proposed \citep{brunetti-blasi05,brunettilazarian11,pinzke17,brunetti17}.

There is broad consensus that radio relics trace shock fronts propagating through the ICM \citep[][for reviews]{roettiger99,ensslin98,pinzke13,bruggen12,brunettijones14}. However, the commonly used diffusive shock acceleration (DSA) model is challenged by the low Mach number of shocks associated with cluster mergers ($\mathcal M\sim1-3$). Therefore the re-acceleration of a population of relativistic electrons might be required to explain the radio brightness of some relics \citep{markevitch05,kang12,vazza14,bonafede14,shimwell15,eckert16,botteon16,vanweeren16,vanweeren17}. 

There is a correlation between the radio power of radio halos and the X-ray luminosity of the hosting clusters, with more powerful radio halos being found in more luminous clusters \citep{liang00,bacchi03,CBS06,rudnick09}. Also, galaxy clusters show a bimodal behaviour in the radio power--X-ray luminosity diagram: radio halos follow the correlation, while upper limits to the radio power of clusters without diffuse emission are well below this scaling relation \citep{brunetti07}. A similar correlation has been found between the radio power of radio halos and the integrated Sunyaev-Zel'dovich (SZ) effect signal \citep{basu12,cassano13,martinezaviles16} as well as the bimodality of clusters in the radio power-SZ diagram connected with the dynamical status of the hosting clusters \citep{cassano13}. Both the X-ray luminosity and the SZ signal are used as proxies of the mass of the clusters, but the SZ signal should be less affected by the dynamical state of the clusters and it is considered to be a more robust indicator of the cluster mass than the X-ray luminosity \citep{motl05,nagai06}. 

\noindent A correlation between the radio power of radio relics and the mass of the host cluster has been recently found, together with the trend of larger relics to be located at larger distance from the cluster center \citep{degasperin14}.

Specifically, for radio halos, the radio bimodality of galaxy clusters and the distribution of clusters in the radio-mass or radio-Lx diagram can be explained in terms of superposition of different evolutionary stages of the cluster life, in connection with their dynamical activity \citep[e.g.][]{brunetti09,cassano16}. In this respect, idealized simulations of -single- two body merger by \citet{donnert13}, which include turbulent re-acceleration of primary electrons, are very instructive. They show that merger-induced turbulence have the potential to light up the radio emission, lifting the clusters from the region of the upper limits in the $P_{1.4GHz}-L_X$ diagram up to the region spanned by the $P_{1.4GHz}-L_X$ correlation. Within $\sim1-2$ Gyr, the radio luminosity is expected to decrease and the clusters go back to the upper limit region. According to these simulations, clusters are expected to spend $\leq0.5$ Gyr in the region between the correlation and the upper limits, thus this region would result statistically less populated \citep{donnert13}. A complication comes from the possibility that secondary particles injected by hadronic collisions may contribute to the cluster-scale radio emission. The level of this contribution is still not understood, yet. It should become dominant in the cases where turbulent acceleration is not efficient enough. This leads to the expectation of ``off state'' halos, produced in less turbulent systems, where the synchrotron emission is maintained by the continuous injection of secondary particles generated from proton-proton collisions \citep[e.g.][]{brunettilazarian11,cassano12}. 
 

In this context, we are studying the occurrence of radio halos in a sample of 75 galaxy clusters with $M_{500}\gtrsim6\times10^{14}M_\odot$ and $z=0.08-0.33$ selected from the Planck SZ cluster catalogue \citep{planck14}. First results, based on a sub-sample of 57 clusters with available radio information, have been published in \citet{cuciti15}.
We are completing the analysis of the sample using new GMRT and JVLA data of all the clusters that lack deep radio observations in order to determine the possible presence of diffuse emission. This work led to the discovery of a number of cluster-scale radio sources. Here we present new GMRT and JVLA data of two clusters, ZwCl 0634.1+4750 (Z0634, hereafter) and Abell 1451 (A1451, hereafter), classified as candidate radio halos in \citet{cuciti15}. Their properties are listed in Table \ref{tab:clusters_prop}.
  
Throughout this paper we assume a $\Lambda$CDM cosmology with $H_0=70$ km s$^{-1}$Mpc$^{-1}$, $\Omega_\Lambda=0.7$ and $\Omega_m=0.3$. With this cosmology 1$''$ corresponds to a physical scale of 2.953 kpc at the redshift of Z0634 and to 3.287 kpc at the redshift of A1451.
\begin{table*}
\centering
\footnotesize
\caption{\label{tab:clusters_prop}Clusters properties}
\begin{tabular}{lccccc}

Name&RA and DEC&$z^*$&$D_L$&$M_{500}^*$&$L_{500}^{**}$[0.1-2.4]kev\\
&(J2000)&&(Mpc)&($10^{14}M_\odot$)&($10^{44}L_\odot$)\\
\hline
\\
Abell 1451& 12 03 16.2 \, $-$21 32 12.7&0.199&978.6&7.32&6.61\\

\\
ZwCl 0634.1+4750&06 38 02.5 \, $+$47 47 23.8&0.174&839.6&6.52&4.72\\
\hline\hline
\end{tabular}
\tablebib{(*)~\citet{planck14}, (**)\citet{MCXC}}
\end{table*}

\section{Radio observations and data analysis}

We performed the data reduction, both calibration and imaging, with CASA \citep[the Common Astronomy Software Applications package,][]{CASA}. An overview of the radio observations presented here is given in Table \ref{tab:clusters_obs}, where we report the name of the clusters, the interferometer and the observing frequency, the date of the observation, the flux calibrator, the phase calibrator, the beam and the rms noise achieved in the images made with the ``Briggs'' weighting scheme (\texttt{robust=0}). Throughout the paper, the error $\Delta_S$ associated with a flux density measurement $S$ is calculated as:
\begin{equation}
\Delta_S=\sqrt{\bigg(rms\times \sqrt{\dfrac{A_s}{A_b}}\bigg)^2+(\sigma_c\times S)^2}
\end{equation}
\label{eq:errori}
where $A_s$ is the source area, $A_b$ is the beam area and $\sigma_c$ is the calibration uncertainty.
Details about the observations and the data reduction are discussed in the next subsections. 

\subsection{GMRT observations}

Z0634 and A1451 have been observed with the GMRT at 320 MHz using an observing bandwidth of 32 MHz subdivided into 256 channels (project code: 28\_066, P.I. V. Cuciti). The total time on source for each cluster is $\sim$5 hours and they have been observed with a duty cycle of 24 min on target and 6 min on the phase calibrator. The flux density calibrators have been observed for $\sim$ 15 min at the beginning and at the end of the observations. 
The flux density scale was set following Scaife \& Held (2012). The bandpass has been corrected using the flux density calibrators. As a first step we obtained amplitude and gain correction for the primary calibrator in 10 central neighbour channels free of RFI, these solutions were applied before determining the bandpass in order to remove possible time variations of the gains during the observation. Once we applied the bandpass, gains solution for all the calibrator sources on the full range of channel were determined and transferred to the target source. Automatic removal of RFI was performed with the CASA task \texttt{flagdata} (\texttt{mode=rflag}). The central 220 channels were averaged to 22 channels each $\sim$1.3 MHz wide to reduce the size of the dataset without introducing significant bandwidth smearing within the primary beam. A number of phase-only self-calibration rounds was carried out to reduce residual phase variations. Wide field imaging was implemented to account for the non-coplanarity of the baselines (wprojection) and wide band imaging (\texttt{mode=mfs}, \texttt{nterms=2}) was used to consider the frequency dependence of the sources in the sky. We applied further manual editing to the data flagging bad self calibration solutions and visually inspecting the residual \textit{uv}-data. However, a number of bright sources in the field of view limited the dynamic range of the image, therefore we obtained direction dependent amplitude and phase solutions for these sources and then we subtracted them out of the \textit{uv}-data. This method is commonly referred to as ``peeling'' technique.

We used the ``Briggs'' weighting scheme with \texttt{robust=0} throughout the self calibration and we produced final ``full resolution'' images with beam $\sim10''$ and rms noise $\sim0.1$ mJy/beam (see Table \ref{tab:clusters_obs} for details). We then produced low resolution images by using different weighting schemes and/or tapering down the long baselines. Images have been corrected for the primary beam response. The residual calibration errors ($\sigma_c$ in Eq. \ref{eq:errori}) are within 10\% \citep[e.g.][]{chandra04}.

\subsection{JVLA observations}

We carried out JVLA L-band observations of Z0634 in D array and in DnC array of A1451 (project code: 15B-035, P.I. V. Cuciti). The hybrid configuration DnC was used to obtain a roundish beam for targets with $\delta<-15^\circ$ such as A1451. The total bandwidth, from 1 to 2 GHz, is divided into 16 spectral windows, each having 64 channels 2 MHz wide. The total time on source for each cluster is $\sim$45 min. We observed the phase calibrators for $\sim$2 min every $\sim$20 min. The flux density calibrators have been observed for $\sim$5 min at the end of the observation.\\
In addition, these two clusters have been observed with the JVLA in L-band in B array as part of a project mainly aimed at studying magnetic fields in galaxy clusters through Faraday rotation measurements of polarized background sources in the Planck ESZ galaxy cluster sample (project 15A-270, P.I. R. van Weeren). The hybrid configuration BnA has been used for A1451, given its southern declination. 
Each cluster has been observed for a total of $\sim$40 min, both 3C286 and 3C147 have been observed as flux density calibrators, while no phase calibrator has been observed in this project. \\
As a first step, the data were Hanning smoothed. We applied the pre-determined antenna position offset and elevation-dependent gain tables. The flux density scale was set according to the Perley-Butler 2013 scale. We determined amplitude and phase solutions for the flux calibrators in the 10 central channels of each spectral window in order to remove possible time variations during the calibrator observation. These solutions were pre-applied to find the delay terms (\texttt{gaintype='K'}) and to correct for the bandpass response. We obtained the complex gain solutions for the calibrator sources (also of the phase calibrator, if present) on the full bandwidth pre-applying the bandpass and delay solutions. Finally, we applied all the calibration tables to the target field. Automatic RFI flagging was applied to the target field using the CASA task \texttt{flagdata} (\texttt{mode=tfcrop}). To reduce the size of the dataset, we averaged the 48 central channels of each spectral window to 6 channels and we averaged in time with a time bin of 15 sec.\\
We ran several rounds of phase-only self calibration on the target field and a final amplitude and phase self calibration. Manual flagging of additional data was applied by visually inspecting the solutions of the self calibration. The wprojection algorithm was used to take into account the non coplanar nature of the array. Wide band imaging is crucial when dealing with the 1 GHz bandwidth of the JVLA, therefore we used at least 2 (sometimes 3) Taylor terms (\texttt{nterms=2-3}) to take the frequency dependence of the brightness distribution into consideration. 
The imaging process involves the use of clean masks that have been made with the \texttt{PyBDSM} package \citep{pybdsm}. For the self calibration we used the ``Briggs'' weighting scheme with \texttt{robust=0} and we made final ``full resolution'' images, whose properties are listed in Table \ref{tab:clusters_obs}. Images have been corrected for the primary beam attenuation. The absolute flux scale uncertainties ($\sigma_c$ in Eq. \ref{eq:errori}) is assumed to be within 2.5\% \citep{perleybutler13}.

After a separate calibration and self-calibration, the B and D array observations were combined together. Unfortunately the pointings did not coincide exactly and this severely affected the quality of the combined image, especially around bright sources away from the phase center. In fact, the primary beam response at the position of those sources is different and this leads to considerable differences in their apparent fluxes between the two observations. To overcome this problem, we subtracted those sources from the \textit{uv}-data and then we combined the datasets. In particular, we performed the subtraction of those sources separately for each spectral window. This is particularly important for sources that are close to the border of the primary beam because the size of the primary beam depends on the frequency, hence the flux density of those sources is considerably different in different spectral windows. The combined B+D array images have a ``full resolution'' of $\sim6 ''$.

\begin{table*}
\centering
\small
\caption{\label{tab:clusters_obs}Radio observations}
\begin{tabular}{lcccccc}
\hline\hline\\
Name& Observation& Date& Flux cal& Phase cal&beam &rms noise \\
&&&&&$(''\times'')$&($\mu$Jy/b)\\
\hline
\\
\bf{Abell 1451}&GMRT 320MHz$^d$&23 May 2015&3C286&1130-148&$11.6\times8.1$&95\\
&JVLA L-band DnC array$^b$&07 Jan 2016&3C286&1130-148&$44.4\times17.1$&60\\
&JVLA L-band BnA array$^c$&18 May 2015&3C147, 3C286&--&$4.7\times3.3$&25\\

\\
\hline
\\
\bf {ZwCl 0634.1+4750}&GMRT 320MHz$^a$&19 Dec 2015&3C147&3C147&$9.7\times7.3$&85\\
&JVLA L-band D array$^b$&19 Oct 2015&3C147&J0713+4349&$35.7\times30.7$&45\\
&JVLA L-band B array$^c$&22 Feb 2015&3C147, 3C286&--&$3.5\times3.3$&26\\
\hline
\end{tabular}
\tablefoot{\textit{a)} obs No 8266 (project code 28\_066), \textit{b)} project code 15B-035, \textit{c)} project code 15A-270, \textit{d)} obs No 7798 (project code 28\_066) .}
\end{table*}

\section{A1451}
\label{Sec:A1451}

A1451 (alternative names: RXC J1203.2-2131, PSZ1 G288.26+39.94) is a massive \citep[$M_{500}=7.32\times10^{14}M_\odot$,][]{planck14} and hot \citep[13.4 keV,][]{valtchanov02} galaxy cluster at redshift $z=0.199$. The MCXC catalogue reports an X-ray luminosity in the $[0.1-2.4]$ keV band $L_{500}=6.61\times10^{44}L_\odot$ \citep{MCXC}. 


Although the X-ray emission of A1451 (Fig. \ref{Fig:A1451_XMM_330} and Fig. \ref{Fig:A1451_XMM_330_rgb}) is fairly regular, several pieces of evidence support the idea of ongoing merging activity for this cluster. Indeed, the high temperature and velocity dispersion are significantly in excess with respect to the estimates derived from weak lensing data \citep{cypriano04}. Moreover, both the X-ray morphology and the mass distribution derived using the distortions of the faint background galaxies \citep{cypriano04} are elongated in the North-South direction. No radio diffuse emission has been detected in A1451 with ATCA 1.4 GHz observations \citep{valtchanov02}. On the other hand, hints of diffuse emission have been detected by \citet{cuciti15} in their re-analysis of the NVSS \textit{uv}-dataset \citep{condon98}.

Here we present the discovery of a faint radio halo at the center of the cluster A1451 and a possible radio relic using a deep GMRT observation at 320 MHz and JVLA observations at 1.5 GHz. These sources are discussed in detail below.

\subsection{Radio halo}
\label{Sec:RH_A1451}
 \begin{figure*}
   \centering
   \includegraphics[width=\hsize,trim={0cm 0cm 0cm 
0cm},clip]{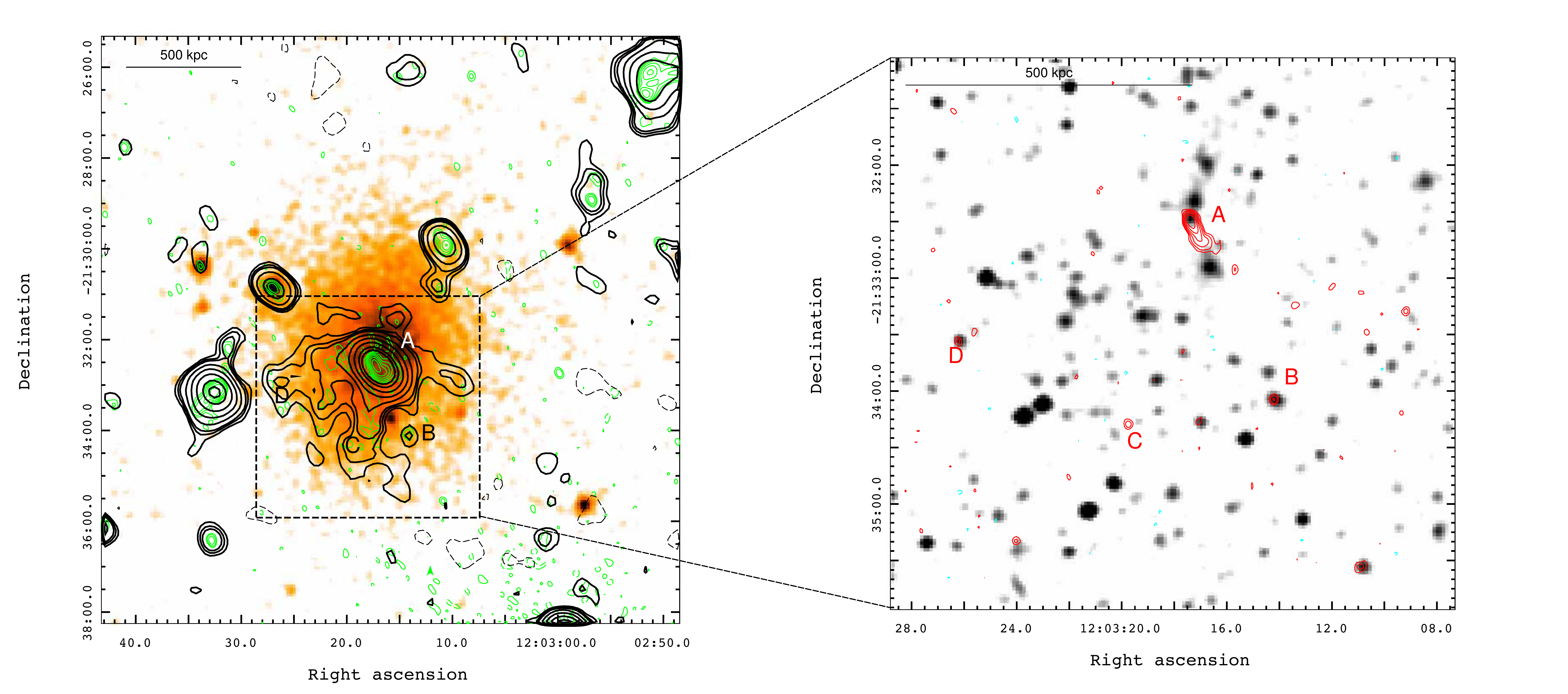}
  
      \caption{\textit{Left}: A1451 GMRT 320 MHz full (green) and low (black) resolution contours superimposed on the X-ray \textit{XMM-Newton} image. Low resolution contour levels (black) are $(2,3,4,8,16,32...)\times \sigma$ with rms noise $\sigma=0.3$ mJy/beam and beam=$39.4''\times27.9''$. Full resolution contours start at 0.3 mJy/beam and are spaced by a factor 2. The full resolution beam is $11.6''\times8.1''$ The first negative contours are dashed. \textit{Right}: JVLA B array contours (red) of the central part of the cluster superimposed on the optical SDSS image. Sources are labelled A to D as in Fig. \ref{Fig:A1451_XMM_330_rgb}. Contours start at 0.08 mJy/ beam and are spaced by a factor 2. The beam is $4.7''\times3.3''$. The first negative contour is shown with a cyan dashed line.
              }
         \label{Fig:A1451_XMM_330}
   \end{figure*}
   
\begin{figure}
   \centering
   \includegraphics[width=\hsize,trim={1cm 0cm 1cm 
0cm},clip]{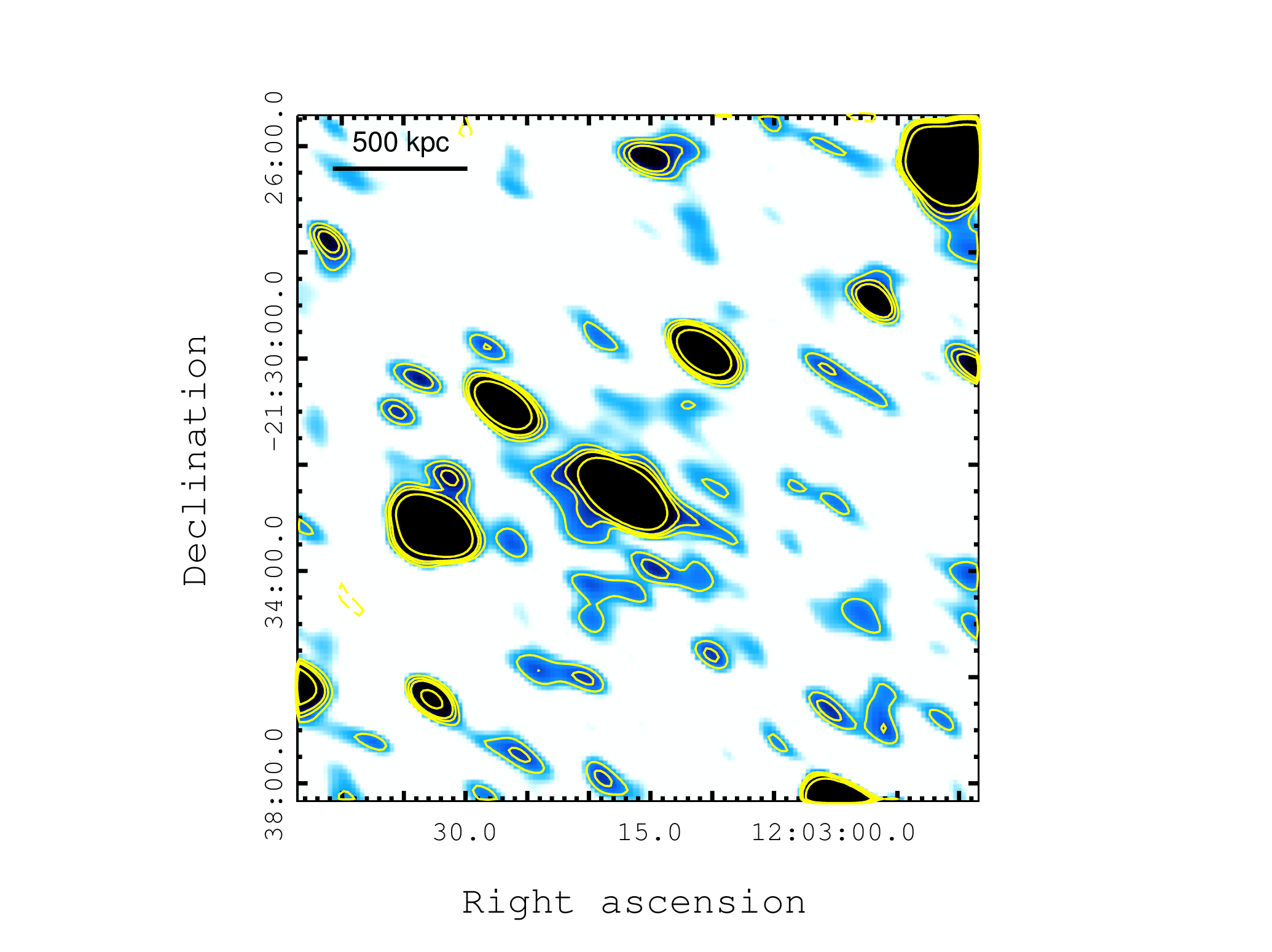}
  
      \caption{JVLA D array colour image and contours of A1451. The contour levels are $(2,3,4,8,16,32...)\times \sigma_{rms}$ with rms noise $\sigma_{rms}=60 \mu$Jy/beam and beam=$44.4''\times17.1''$. The first negative contour is dashed.
              }
         \label{Fig:A1451_D}
   \end{figure}
A number of discrete radio sources are visible on the cluster area in the GMRT 320 MHz full resolution image (labelled A to D in Fig. \ref{Fig:A1451_XMM_330}). In particular, there is a radio galaxy at the center of the cluster (source A), with a total flux density of $\sim80$ mJy at 320 MHz and $\sim17$ mJy at 1.5 GHz. The JVLA B array image (Fig. \ref{Fig:A1451_XMM_330}, right panel) shows that the brighter part of this source is followed by a short, low surface brightness tail that makes its subtraction from the \textit{uv}-data difficult. 
In addition to the individual sources, some positive residuals are visible in the full resolution GMRT 320 MHz image (green contours in Fig. \ref{Fig:A1451_XMM_330}, left panel). The low resolution image (black contours in Fig. \ref{Fig:A1451_XMM_330}, left panel) clearly shows the presence of extended emission, that we classify as a radio halo. The radio halo has a largest angular size (LAS) of $\sim4$ arcmin in the N-S direction, corresponding to a projected largest linear size (LLS) of $\sim$750 kpc. It is mainly extended toward the Southern part of the cluster. There are hints of diffuse emission also in the Northern part, both in the GMRT and JVLA images (Fig.\ref{Fig:A1451_D}), but their surface brightness is below twice the rms noise level of the images.
We measured the flux density of the radio halo on the low resolution image (Fig. \ref{Fig:A1451_XMM_330}, left panel), then we subtracted the flux density of the central radio galaxy and of the three point-like sources. Specifically, we measured the flux density of the point-like sources on the full resolution ($\sim10''$) image, while we measured the flux density of source A on the low resolution image, to take into account also its extended emission. We obtained a flux density for the radio halo of $S_{320MHz}\sim22$ mJy. It is likely that subtracting source A we are also subtracting part of the radio halo flux density. We measured the mean surface brightness of the central part of the radio halo (around source A) and we estimated that in the area of source A there are $\sim$5 mJy of radio halo flux density, leading to a total flux density of the radio halo $S_{320MHz}= 27.3\pm2.3$ mJy.
 
An equivalent procedure has been applied to measure the flux density of the radio halo at 1.5 GHz, i.e. we measured the radio flux density on the whole region of the halo from the D array image (Fig. \ref{Fig:A1451_D}, resolution$\sim40''$) and we subtracted the flux density of the central source, measured on the same map, and of the three point-like sources, taken from the B array image (resolution$\sim4''$). The flux density of the radio halo is $S_{1.5GHz}=5.0\pm0.5$ mJy, which corresponds to a radio power $P_{1.4GHz}=6.4\pm0.7\times10^{23}$ W/Hz, assuming a spectral index $\alpha=-1.2$ for the \textit{k-correction}. These values take into account the portion of the radio halo covered by source A.

To evaluate the spectrum of the halo between 320 and 1500 MHz (B+D array JVLA combined observations) we produced images with uniform weighting, using the common \textit{uv}-range and tapering down to the same resolution ($\sim20''$) and we measured the flux density of the radio halo exactly on the same region.
We performed the imaging with two approaches: \textit{i)} cleaning only inside the mask containing all the sources in the field until the residuals are $\sim$two times the rms noise of the image or \textit{ii)} after that, cleaning also on the rest of the field down to two times the rms noise. Depending on how we cleaned the images and on the region where we measure the flux density of the radio halo we found values for the spectral index in the range $-1.3\lesssim\alpha\lesssim-1.1$. In particular, we found the steepest spectral indices in case \textit{(ii)}. We note that a higher cleaning threshold does not yield different results, indeed, if we clean until the residuals are $\sim$3 times the rms noise of the images, we obtain no significant difference in the flux density of the radio halo at both frequencies and in both cases \textit{(i)} and \textit{(ii)}.

   \begin{figure}
   \centering
   \includegraphics[scale=0.45,trim={1.5cm 1cm 1cm 
0cm},clip]{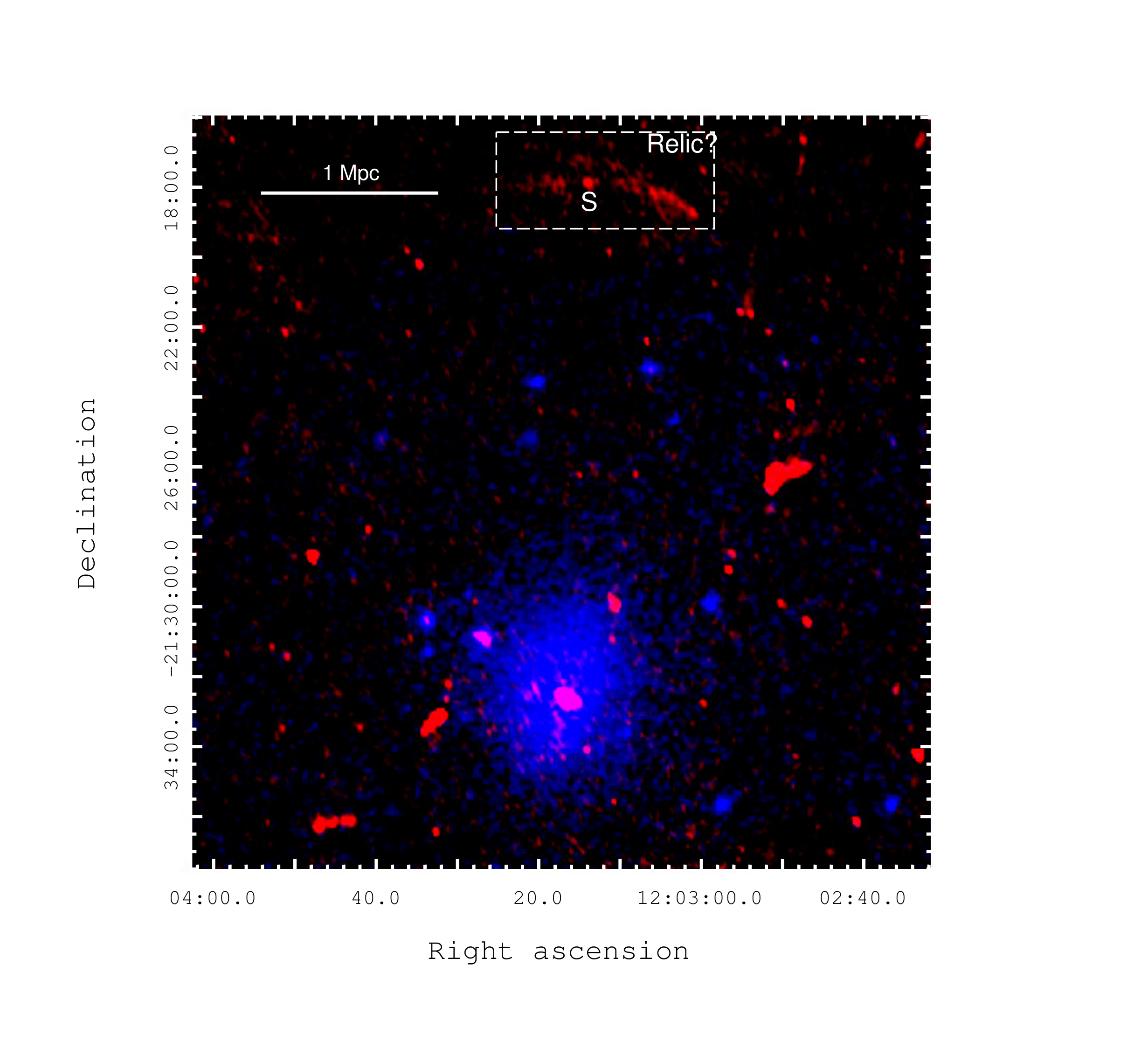}
      \caption{A1451 GMRT 320 MHz image (red) superimposed on the X-ray \textit{XMM-Newton} image (blue). 
              }
         \label{Fig:A1451_XMM_330_rgb}
   \end{figure}
\subsection{A distant relic?}
\label{Sec:relic}

\begin{figure*}
   \centering
   \includegraphics[width=\hsize,trim={0cm 3cm 0cm 
0cm},clip]{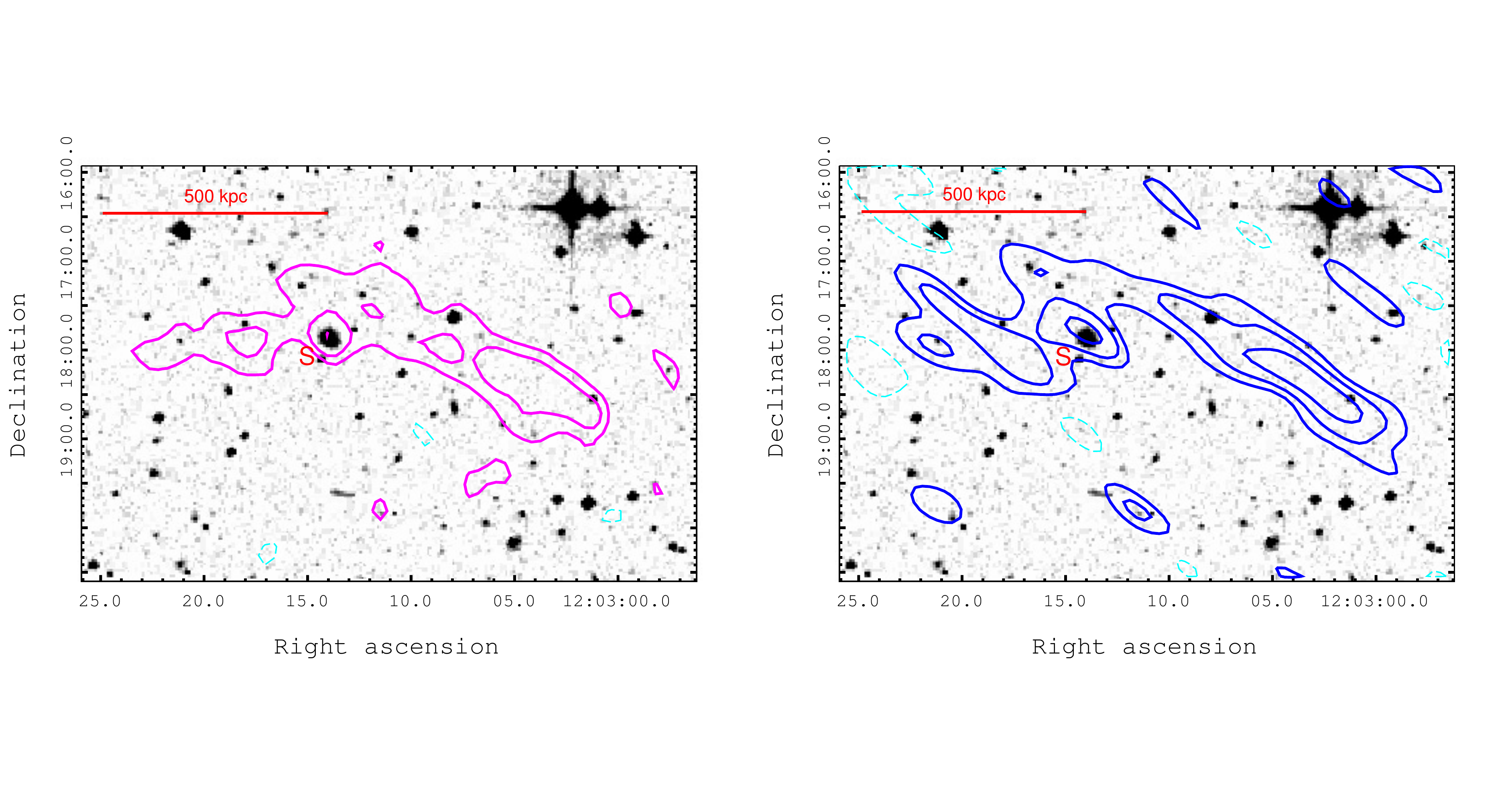}
      \caption{GMRT 320 MHz (\textit{left}) and JVLA D array 1.5 GHz (\textit{right}) contours of the relic in A1451 superimposed on the optical SAO-DSS image. \textit{Left}: radio contours starts at 0.6 mJy/beam and are spaced by a factor 2. The first negative contour is shown with a cyan dashed line. The rms noise of the radio image is $\sim0.2$ mJy/beam with a beam of $23.3''\times18.0''$. \textit{Right}: radio contours starts at 0.18 mJy/beam and are spaced by a factor 2. The first negative contour is shown with a cyan dashed line. The rms noise of the radio image is $\sim60 \mu$Jy/beam with a beam of $44.4''\times17.1''$.
              }
         \label{Fig:A1451_relic}
   \end{figure*}
   
We detected an elongated diffuse source located $\sim15$ arcmin North of the galaxy cluster center (Fig. \ref{Fig:A1451_XMM_330_rgb}), both at 320 MHz and in the JVLA D array 1.5 GHz image (Fig. \ref{Fig:A1451_relic}). Based on its morphology and location we classify it as a candidate radio relic. A zoom on the candidate relic, superimposed on the optical SAO-SDSS image, is shown in Fig. \ref{Fig:A1451_relic}. If assumed to be at the redshift of the cluster, this source is $\sim3$ Mpc away from the cluster center. Its LAS in the E-W direction is $\sim6.5$ arcmin, which would correspond to a LLS$\sim1.3$ Mpc. A point-like radio source with a clear optical counterpart (labelled S in Fig. \ref{Fig:A1451_relic}) is embedded in the diffuse emission. With the same procedure used for the radio halo, we measured the flux density of the point-like source from high resolution images and we subtracted it from the flux density measured on the candidate relic area, obtaining a flux density of $35.5\pm5.3$ mJy at 320 MHz and $9.0\pm0.5$ mJy at 1.5 GHz.  

To evaluate the spectral index of the candidate radio relic between 320 MHz and 1.5 GHz (D array observation), we used uniform weighting, the common \textit{uv}-range and we convolved the images to the same resolution ($50''$). The integrated spectral index of this source is $\alpha=-1.1\pm0.1$. The resolution of these observations does not allow a reliable analysis of a possible spectral index gradient along the N-S axis. Conversely, the data allow us to conclude that there is no evidence of a spectral index gradient along the E-W axis, as expected in case of an active source in S refurbishing the extended emission. The radio power of the candidate relic at 1.4 GHz is $P_{1.4GHz}=1.13\pm0.06\times10^{24}$ W/Hz, assuming a spectral index $\alpha=-1.1$ for the \textit{k-correction}.

\section{Zwcl0634.1+4747}
\label{Sec:Z0634}
Little is known in the literature about Zwcl0634.1+4747 (alternative names: PSZ1 G167.64+17.63, CIZA J0638.1+4747, RX J0638.1+4747). It is at redshift $z=0.174$ and the MCXC catalogue reports an X-ray luminosity in the $[0.1-2.4]$ keV band $L_{500}=4.72\times10^{44}L_\odot$ \citep{MCXC}. The mass within $R_{500}$ estimated by \cite{planck14} is $M_{500}=6.52\times10^{14}M_\odot$. X-ray \textit{Chandra} archival observation of Z0634 has been analysed in \citet{cuciti15} where evidence of a non relaxed dynamical state has been found. Indeed the \textit{Chandra} image (Fig. \ref{Fig:Z0634_Chandra_D}, left panel) shows the presence of substructures in the X-ray surface brightness distribution and the morphology of the cluster is elongated in the East-West direction.

The JVLA D array image of Z0634 is shown in Fig. \ref{Fig:Z0634_Chandra_D} (left panel). The cluster hosts diffuse, centrally located radio emission, that we classify as a radio halo. Two head-tail radio galaxies are present in the cluster field, one located North of the radio halo (labelled HT in Fig. \ref{Fig:Z0634_Chandra_D}) and one located at the S-E edge of the cluster (source C in Fig. \ref{Fig:Z0634_Chandra_D}). While source C has the typical morphology and spectral properties of many head tail radio galaxies found in clusters \citep[e.g.][]{o'dea&owen86,giacintucci09,owen14}, the radio galaxy named HT has some unusual characteristics. The radio halo and the HT will be described in the following subsections.

\subsection{Radio halo}
\label{Sec:RH_Z0634}
\begin{figure*}
   \centering
   \includegraphics[width=\hsize,trim={0cm 0cm 0cm 
0cm},clip]{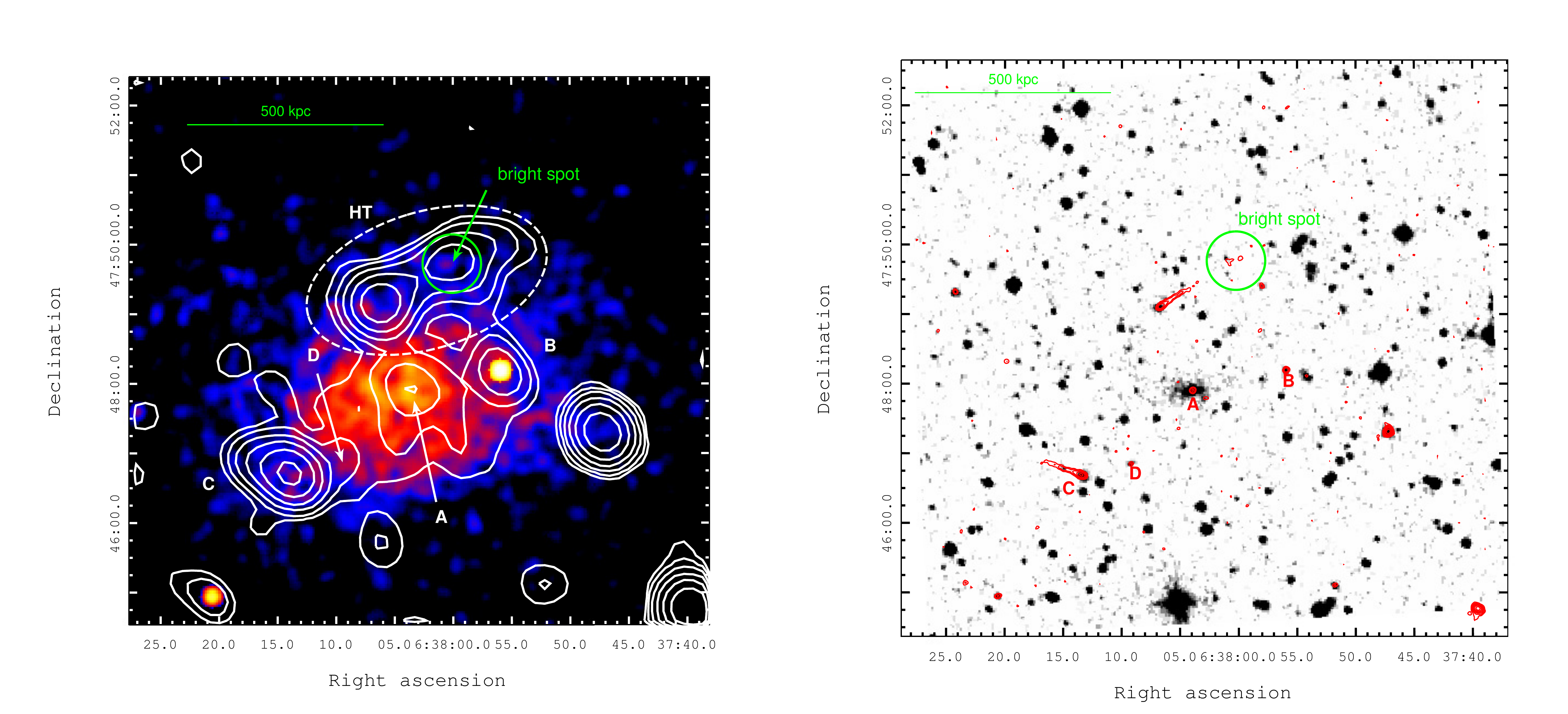}
      \caption{\textit{Left}: Z0634 JVLA D array 1.5 GHz contours (white) superimposed on the X-ray \textit{Chandra} image. Radio contours starts at 0.12 mJy/beam and are spaced by a factor 2. The rms noise of the radio image is $\sim45 \mu$Jy/beam with a beam of $35.7''\times30.7''$. \textit{Right}: JVLA B array contours (red) superimposed on the optical SDSS image. Contours starts at 0.09 mJy/beam and are spaced by a factor 2. The beam is $3.5''\times3.3''$.
              }
         \label{Fig:Z0634_Chandra_D}
   \end{figure*}

The radio halo in Z0634 extends over $\sim600$ kpc in the E-W direction and follows the morphology of the X-ray emission of the cluster (Fig. \ref{Fig:Z0634_Chandra_D}, left panel). The two head-tail radio galaxies, HT and C, are difficult to subtract from the \textit{uv}-data, thus we measured the radio halo flux density on an area that avoids these two sources. Still, there are three sources (labelled A, B and D in Fig. \ref{Fig:Z0634_Chandra_D}) embedded in the radio halo emission. The high resolution ($\sim3''$) JVLA B array image (Fig. \ref{Fig:Z0634_Chandra_D}, right panel) reveals that they are point-like sources, so their flux density, measured from the B array image ($\sim1.4$ mJy in total), can be safely subtracted out from the measure on the radio halo region. We obtained a flux density of the radio halo $S_{1.5GHz}=3.3\pm0.2$ mJy, corresponding to a radio power $P_{1.4GHz}=3.1\pm0.2\times10^{23}$ W/Hz, assuming a spectral index $\alpha=-1.2$ for the \textit{k-correction}.

\begin{figure}
   \centering
   \includegraphics[width=\hsize,trim={1cm 0cm 1cm 
2cm},clip]{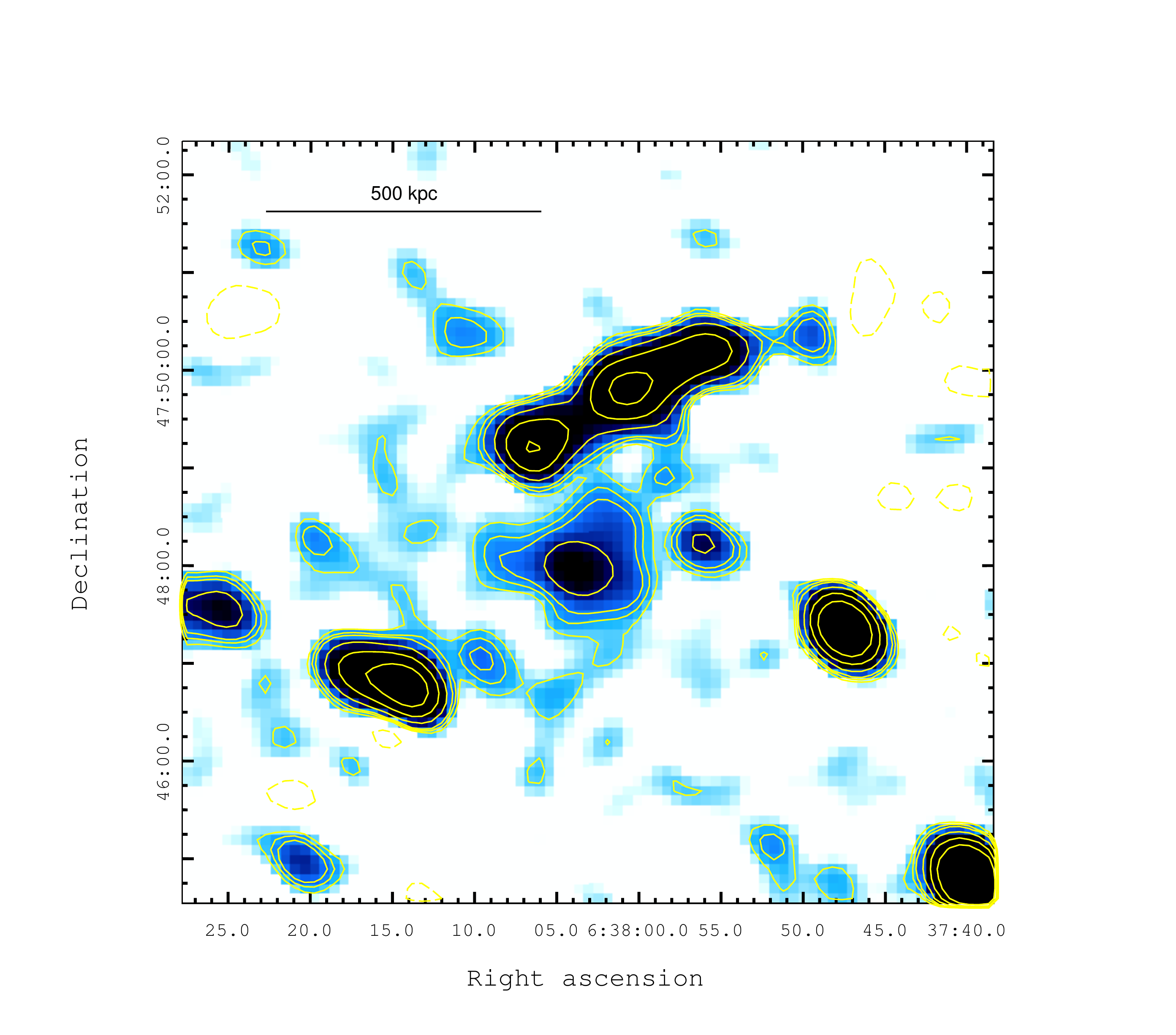}
  
      \caption{GMRT 330 MHz colour image and contours of Z0634. The contour levels are $(2,3,4,8,16,32...)\times \sigma_{rms}$ with $\sigma_{rms}=$ 0.25 mJy/beam. The resolution is $30.5''\times23.0''$. The first negative contour is dashed.
              }
         \label{Fig:Z0634_330}
   \end{figure}

The flux density of the radio halo at 320 MHz, after the subtraction of the point sources mentioned above, is $S_{320MHz}= 20.3\pm2.7$ mJy. We note that in the low resolution GMRT 320 MHz image (Fig. \ref{Fig:Z0634_330}) the radio halo does not extend to the HT and source C, thus we can reasonably exclude that we have lost a substantial part of the radio halo emission by avoiding those regions in the measurement of the radio halo flux density. To evaluate the spectral index of the radio halo, we produced images at 320 and 1500 MHz (B+D array JVLA observations) with uniform weighting and using the same \textit{uv}-range and resolution ($\sim20''$) and we measured the flux density of the radio halo on the same area. As done for A1451, we performed the imaging with two approaches: \textit{i)} cleaning only inside the mask that contains all the sources in the field or \textit{ii)} after that, cleaning also on the rest of the field down to two times the rms noise. We found that the spectral index of the radio halo ranges from --1 to --1.3, depending on where we measure the flux densities and how we clean the images ($\alpha$ is steeper in case \textit{(ii)}). We obtained similar results setting the threshold at 3 times the rms noise in the cleaning process.

\subsection{Head tail}

\begin{figure*}
   \centering
   \includegraphics[width=\hsize,trim={1cm 2cm 1cm 
2cm},clip]{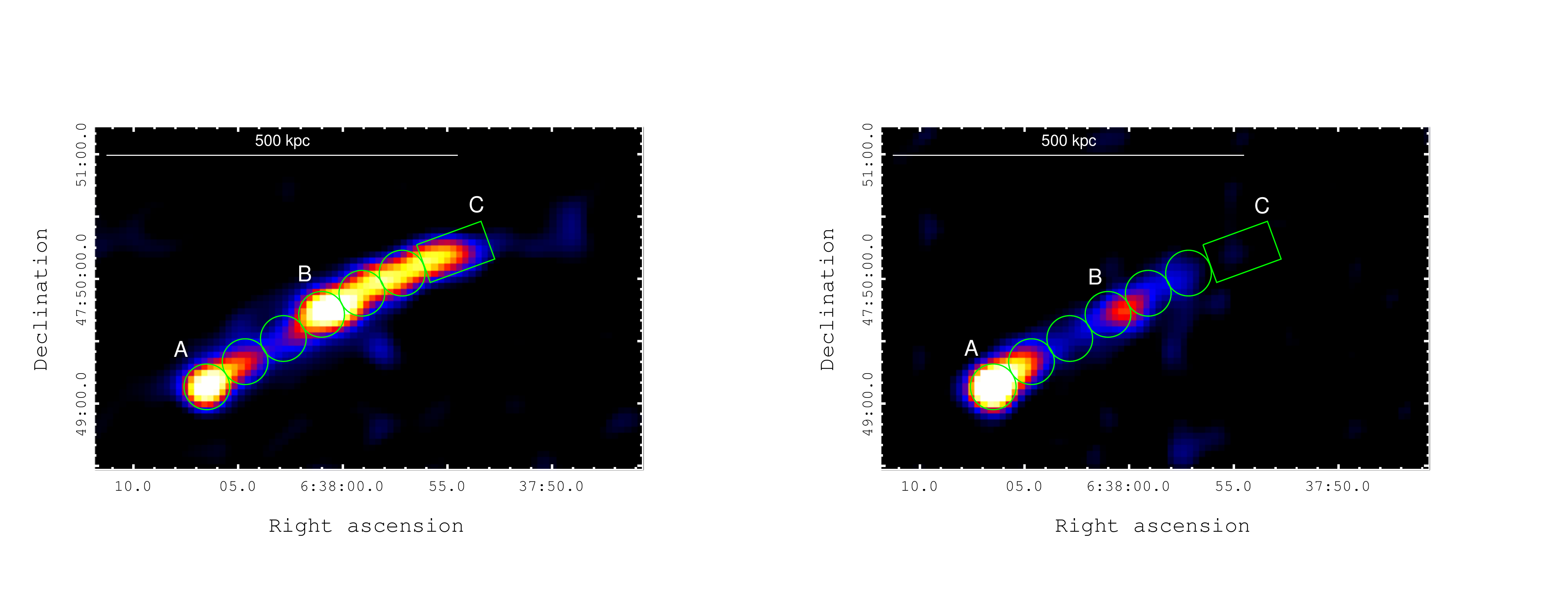}
      \caption{GMRT 320 MHz (\textit{left}) and JVLA B+D array 1.5 GHz (\textit{right}) colour images. Both are convolved with a $15''\times15''$ beam. The green regions mark the areas where we calculated the spectral index to trace its evolution along the tail.
              }
\label{Fig:coda_profile}
   \end{figure*}

\begin{figure*}
   \centering
   \includegraphics[scale=0.5,trim={0.5cm 3cm 0.1cm 
0cm},clip]{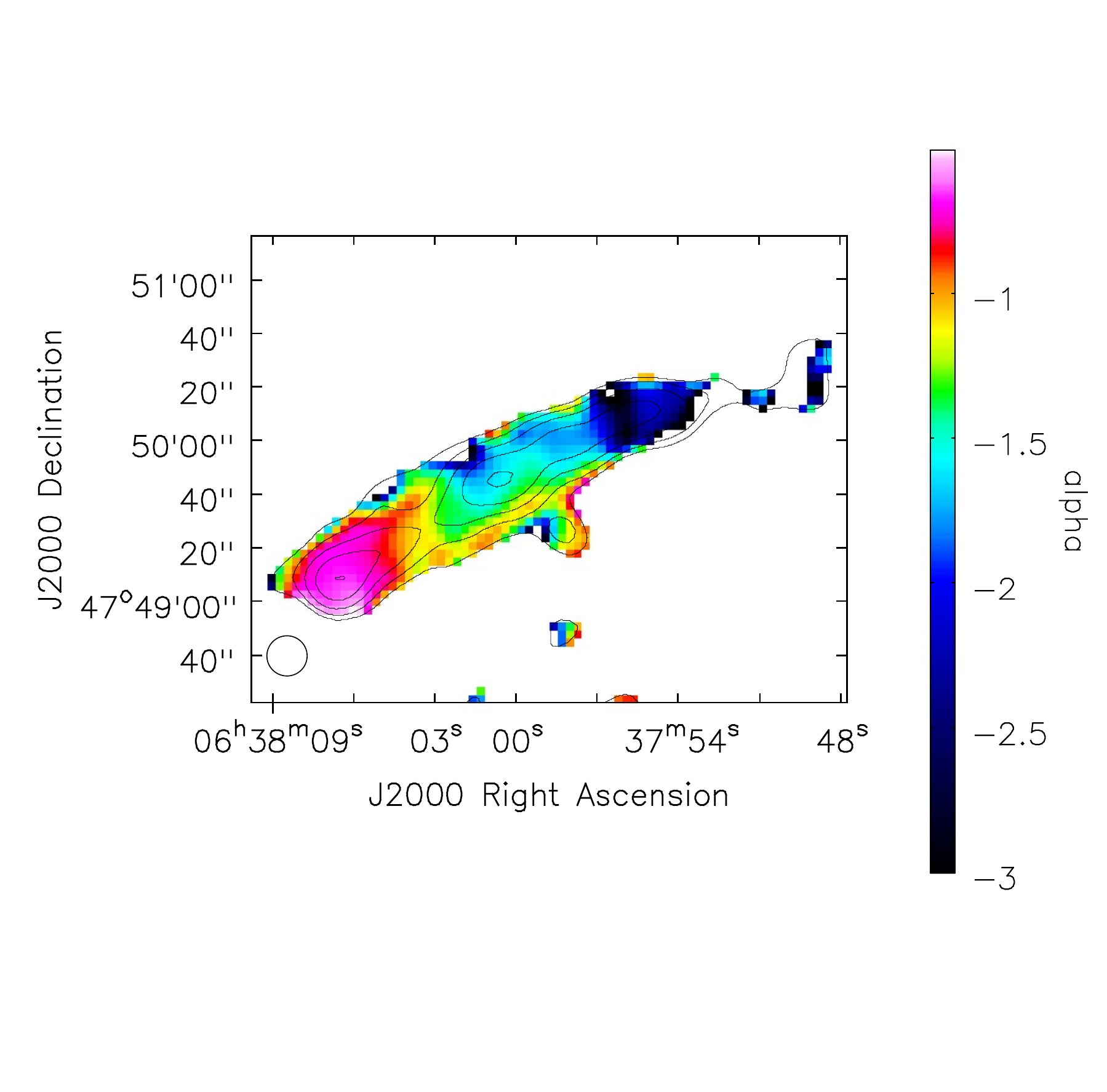}
\includegraphics[scale=0.5,trim={0.5cm 3cm 0.1cm 
0cm},clip]{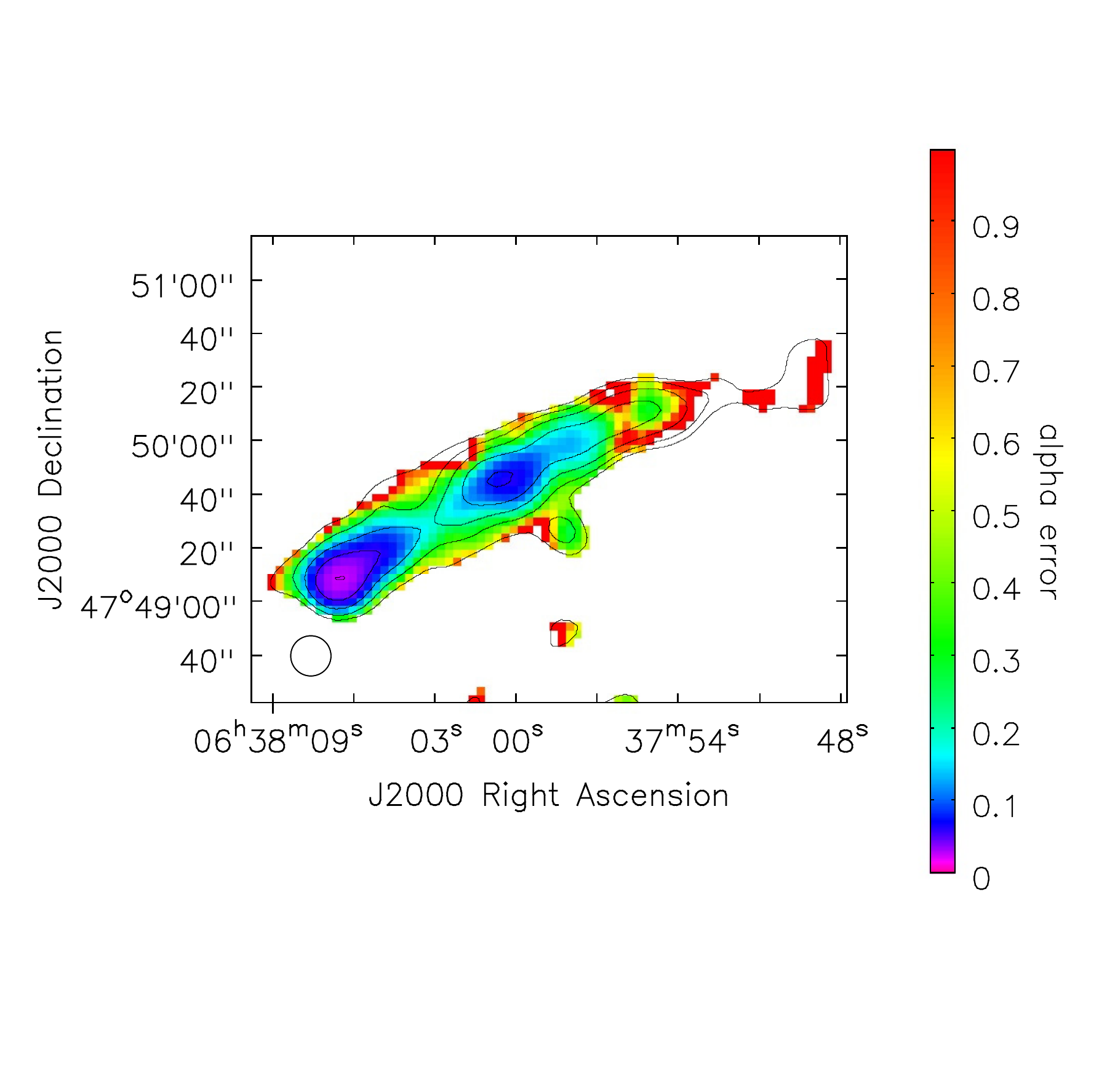}
      \caption{Spectral index map (\textit{left}) and associated spectral index error map (\textit{right}) of the HT in Z0634 between 320 MHz and 1.5 GHz at resolution $15''\times15''$. GMRT 320 MHZ contours of the image used to make the spectral index map are drawn in black. Contours levels start from 0.42 mJy/beam and are spaced by a factor 2.
              }
         \label{Fig:Z0634_codaspix}
   \end{figure*}

The ``head'' of the head tail radio galaxy is located $\sim110''$ NNE from the cluster X-ray centroid (i.e. at a projected distance of $\sim330$ kpc). This bright nucleus is followed by a lower surface brightness tail that extends in the SE-NW direction (Fig.\ref{Fig:Z0634_Chandra_D}). Considering the 3$\sigma$ contours of the GMRT 320 MHz image (shown in Fig. \ref{Fig:Z0634_codaspix}), the angular extension of the HT is $\sim3.5$ arcmin, corresponding to $\sim600$ kpc at the cluster redshift (Fig.\ref{Fig:coda_profile}, left panel). Its linear extension decreases to $\sim400$ kpc at 1.5 GHz (Fig.\ref{Fig:coda_profile}, right panel), measured from the 3$\sigma$ contours of the B+D array image. The total flux density of the HT at 320 MHz is $42.5\pm4.3$ mJy, while it is $8.9\pm0.3$ mJy at 1.5 GHz. The radio surface brightness does not decrease regularly along the tail. There is a spot of higher surface brightness, roughly at half length, visible at both frequencies. After this spot, on the Western side, the tail is brighter than before the spot (Fig.\ref{Fig:coda_profile}). 

To make the spectral index map of the HT we used the spectral windows 6 and 7 of the JVLA B+D dataset and we produced JVLA and GMRT images with uniform weighting and common \textit{uv}-range and we convolved them to the same resolution ($15''$). We blanked those pixels that in the GMRT 320 MHz image have values below two times the rms noise ($\sim0.25$ mJy/beam), in order to take into account the extension of the tail at low frequency as much as possible. However, given the poor SNR and the large errors associated with the spectral index (Fig. \ref{Fig:Z0634_codaspix}), in the analysis in Sect. \ref{Sect:HT}, we will not consider the westernmost part of the tail. 
As shown in Fig. \ref{Fig:Z0634_codaspix}, we found that the spectral index between 320 MHz and 1.5 GHz steepens with the distance from the nucleus, from $\alpha\sim-0.6$ to $\alpha\sim-2.5$. We also note that the bright spot is steeper that the Eastern (initial) part of the tail.

\begin{table*}
\centering
\caption{Properties of the extended sources\label{tab:obs_results}}
\begin{tabular}{lcccccc}
\hline\hline\\
Name& Detection & $S_{320 MHz}$& $S_{1.5 GHz}$&$P_{1.4GHz}$&Integrated $\alpha$&LAS\\
&&(mJy)&(mJy)&($10^{23}$ W/Hz)&&($'$)\\
\hline\\

\multirow{2}{*}{A1451}&RH&$27.3\pm2.3$&$5.0\pm0.6$ &$6.4\pm0.7$&$>-1.3^*$&4.2\\
    &RR&$35.5\pm5.3$&$9.0\pm0.5$&$11.3\pm0.6$&$-1.1\pm0.1$&6.5\\
    \hline
    \\
    
\multirow{2}{*}{Z0634}&RH&$20.3\pm2.7$&$3.3\pm0.2$&$3.1\pm0.2$& $>-1.3^{**}$&3.3\\
    &HT&$42.5\pm4.3$&$8.9\pm0.3$&$8.5\pm0.3$&$-1.25\pm0.08$&3.5\\
    \hline
\end{tabular}
   \tablefoot{Col. 1: cluster name; Col. 2: detected source; Col. 3: flux density at 320 MHz; Col. 4: flux density at 1.5 GHz; Col. 5 radio power at 1.4 GHz; Col. 5: integrated spectral index between 320 MHz and 1.5 GHz. * See Sect. \ref{Sec:RH_A1451} and \ref{Sec:test}. ** See Sect. \ref{Sec:RH_A1451} and \ref{Sec:test}.} 
 \end{table*}

\section{Discussion} 

\begin{figure}
   \centering
   \includegraphics[scale=0.4,trim={0cm 4cm 0cm 
4cm},clip]{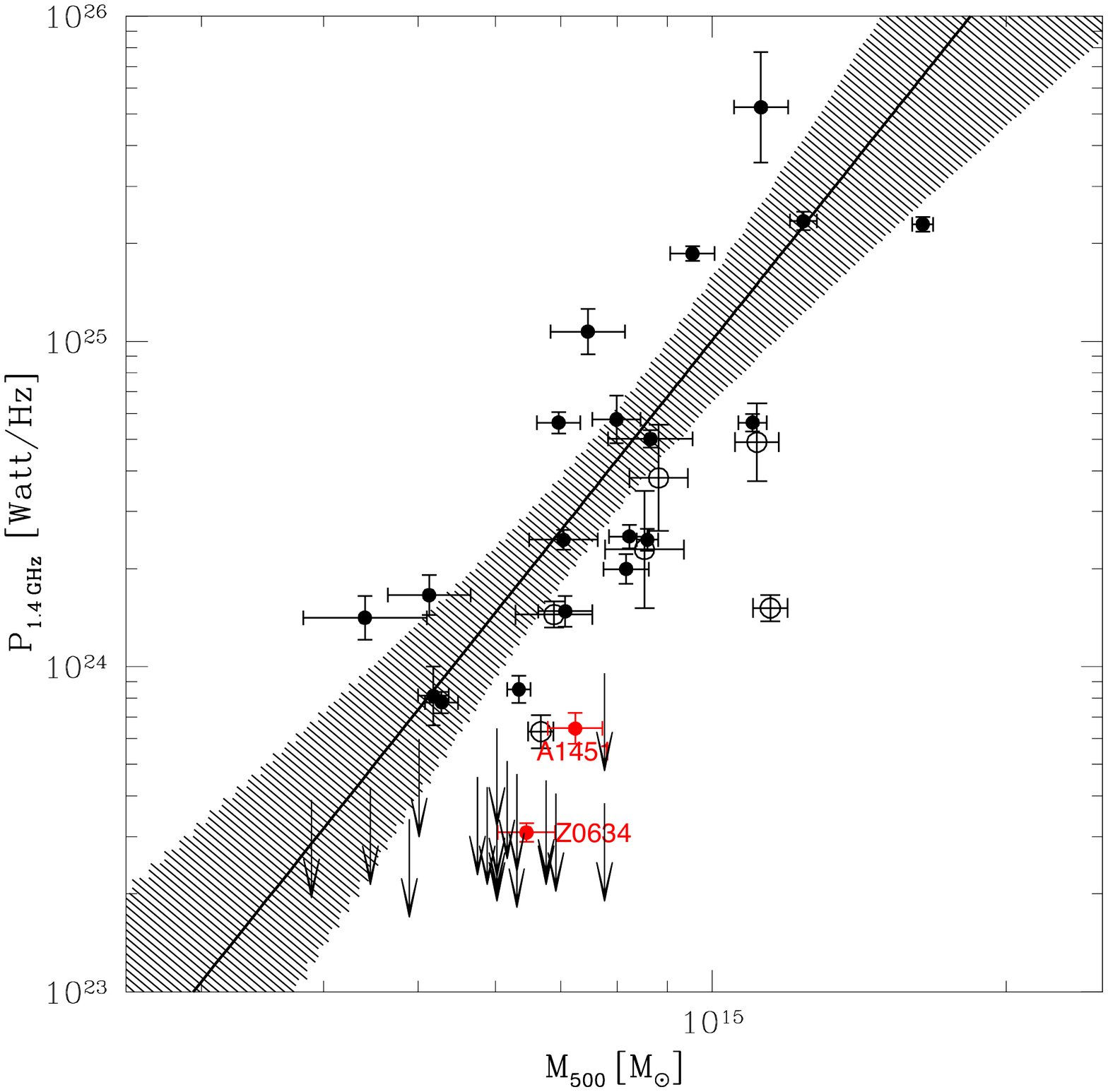}
      \caption{Distribution of clusters in the $P_{1.4GHz}-M_{500}$ diagram. Different symbols indicate giant radio halos (black filled dots), ultra steep spectrum radio halos (black empty dots) and upper limits (black arrows) belonging to the Cassano et al. 2013 sample. The red dots represent the two clusters presented in this paper. The best fit relation for giant radio halos (black line) and the 95\% confidence level (shadowed region) from \citet{cassano13} are shown.
              }
         \label{Fig:P-M}
   \end{figure}
   
\subsection{Testing the reliability of the spectral indices of the radio halos}
\label{Sec:test}
Compared to the current radio power--mass scaling relations \citep{cassano13,martinezaviles16}, the radio halos in A1451 and Z0634 are underluminous, being $\sim6-7$ times below the correlation (Fig. \ref{Fig:P-M}). 
Since these radio halos are very faint, a reliable characterization of their morphology and spectrum is challenging with the current observations.
Here we investigate the possibility that the measure of these spectral indices is biased because of the different capability of our GMRT and JVLA observations to properly recover such extended faint radio emission. 
For this purpose, we used the method of injecting fake radio halos in the datasets. This technique has been used in the literature to place upper limits to the diffuse radio flux density of clusters in which a radio halo is not detected \citep{brunetti07,venturi08,russell11,kale13,kale15}. 
In fact, an important outcome of those studies is that the flux of the fake radio halos is at least partly lost and this effect becomes more important for faint flux densities \citep[e.g.][]{venturi08}. 
Therefore this warns that part of the flux of the radio halos in A1451 and Z0634 may not be properly recovered by our GMRT and JVLA observations. 


We modelled the radio halo brightness profile with an exponential law in the form:
\begin{equation}\label{eq:model_halo}
I(r)=I_0 e^{-\frac{r}{r_e}}
\end{equation}
where $I_0$ is the central surface brightness and $r_e$ is the $e$-folding radius \citep{orru07,murgia09}. From the radio power--radio halo radius correlation \citep[equation 6 in][]{cassano07} we calculated the reference radii of the radio halos, $r_h$, assuming that they follow the radio power--mass correlation \citep[equation 14 in][]{cassano13}. At first, we used $r_e=r_h/2.6$, where 2.6 is the median value of the quantity $r_h/r_e$ for the radio halos studied by both \citet{murgia09} and \citet{cassano07}, then we slightly change the value of $r_e$ in the injected halos to reproduce the extension of the halos observed at the cluster center. In the injected models, the radio halo profile was set to zero for $r>5r_e$. 
  
For each cluster, we choose a region in the image void of bright sources and clear noise pattern and we created a set of fake radio halos with different integrated flux densities, centred on that region. We started injecting a fake radio halo that would lie on the radio power--mass correlation \citep{cassano13} and we reduced the injected flux density until the recovered flux of the fake radio halo was similar to that of the radio halo that we observe at the cluster center. Each model has been Fourier transformed into the MODEL\_DATA column of the dataset including the wprojection algorithm to take into account the large field of view. 
We added the fake radio halo to the original visibilities and we imaged the modified dataset. We measured the recovered flux density of the fake radio halo and we compared it with the injected flux. Since the main aim of these injections is to check the reliability of the measured spectral indices of the two radio halos presented here, the images have been done minimising the differences between the two datasets as much as possible. 
In particular, we used the common \textit{uv}-range and we tapered down to the same resolution ($\sim20''$), as done in Sect. \ref{Sec:RH_A1451} and \ref{Sec:RH_Z0634}. We also used uniform weighting to minimize the difference in the \textit{uv}-sampling. 
We performed the imaging with two approaches: \textit{(i)} cleaning only inside the mask containing all the sources or \textit{(ii)} cleaning deeply also on the rest of the field. In general, we found that the rms noise of the final image improves in case \textit{ii)}, but the recovered flux of the injected radio halos is significantly lower (from 10 to 40\%) than in case \textit{i)}, such effect being more prominent for faint injected flux densities.

\begin{figure*}
   \centering
   \includegraphics[scale=0.4]{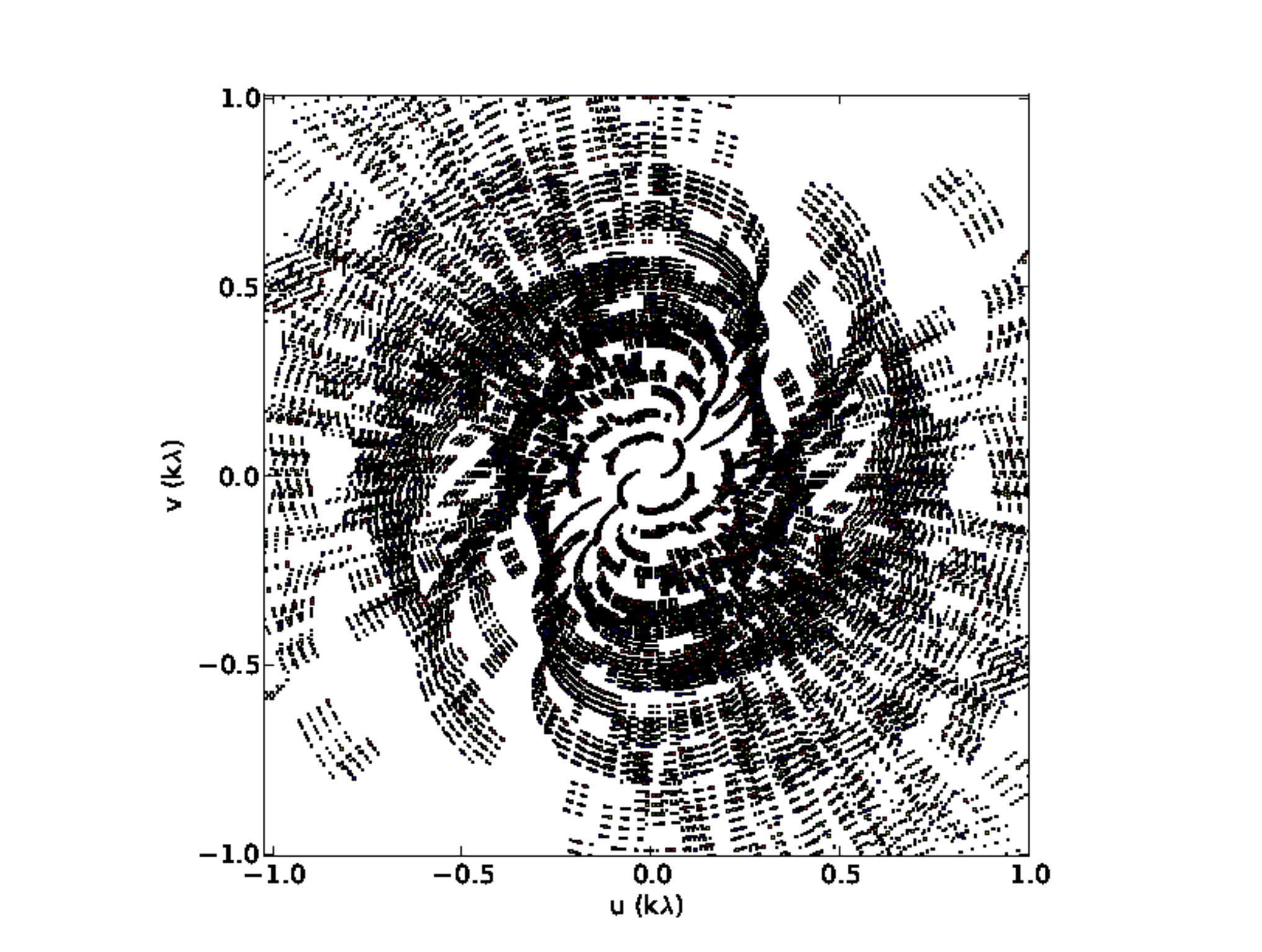}
	\includegraphics[scale=0.4]{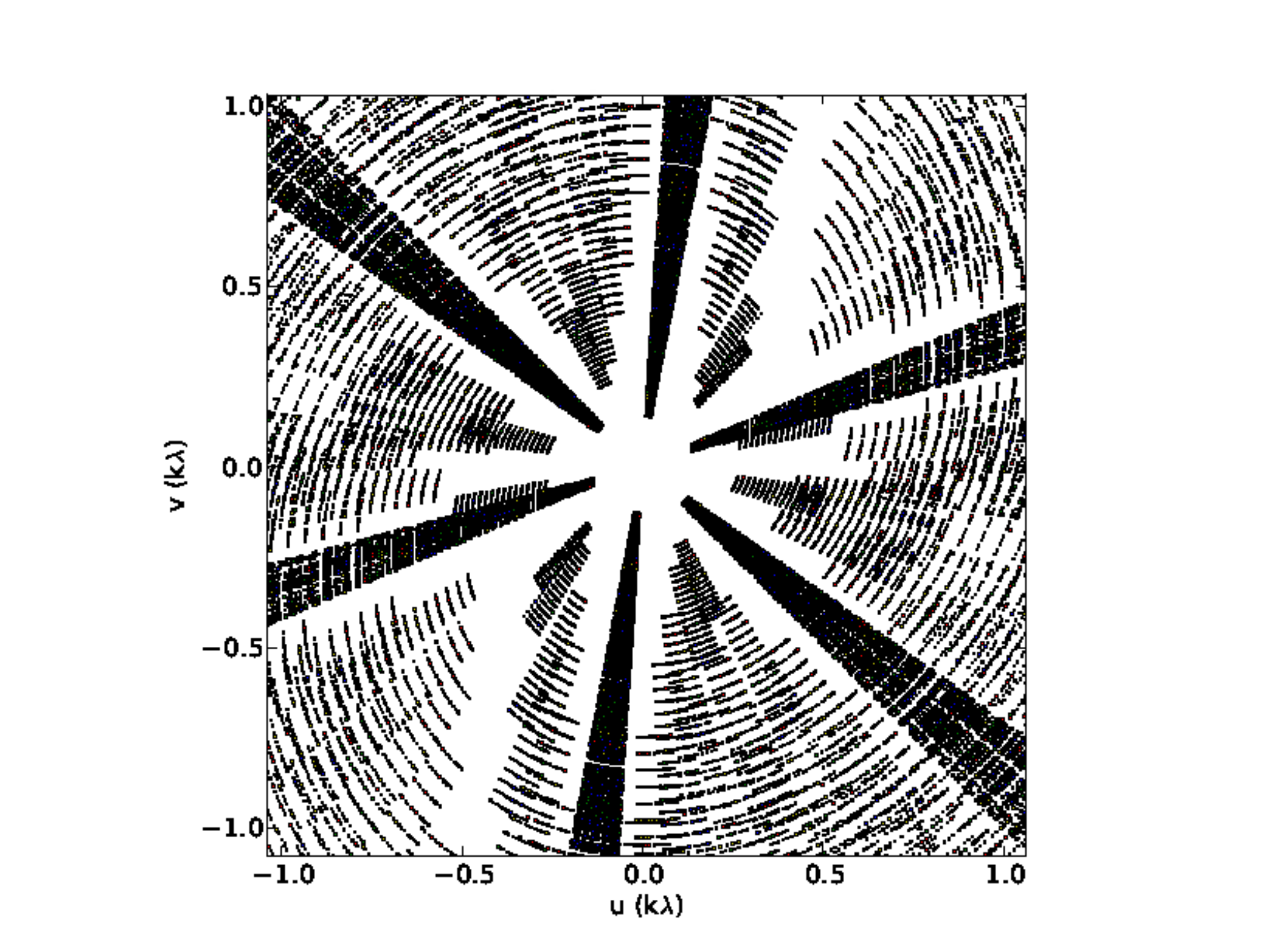}
      \caption{Inner \textit{uv}-coverage of the GMRT 320 MHz observation (\textit{left}) and of the JVLA D array observation (\textit{right}) of Z0634.
              }
         \label{Fig:Z0634_uvcov}
   \end{figure*}

An important outcome of this analysis is that, for both clusters, if we inject the flux density of a radio halo that lies on the radio luminosity-mass correlation, both interferometers are able to recover the great majority of its flux density. However, reducing the injected flux, in order to match the observed one, the fraction that is recovered decreases. The decline of the recovered flux depends on a combination of sensitivity of the observations, density of the inner uv coverage and image fidelity of the dataset \citep{venturi08,kale13,kale15}. More specifically, we found that our GMRT observations generally recover an higher fraction of the flux density of faint injected radio halos, assuming a spectral index $\alpha\leq-1.2$. This different performance can be explained by the different inner \textit{uv}-coverage. Indeed, the full-track GMRT observation ensures a denser \textit{uv}-sampling with respect to the snapshot JVLA observation (Fig. \ref{Fig:Z0634_uvcov}). For both A1451 and Z0634, in order to recover in the fake halo the same flux density that we find in the halo at the cluster center at both frequencies, we need to inject a radio halo with a spectral index $\alpha\approx-1$. Unfortunately, a fully reliable estimate of the spectral index of these radio halos is prevented by the fact that the flux density of such faint diffuse emission can change significantly depending on where we measure it and how we perform the imaging (see \ref{Sec:RH_A1451} and \ref{Sec:RH_Z0634}). Nevertheless, the analysis based on the injections allows us to exclude the presence of radio halos with spectra significantly steeper than $\alpha\approx-1.3$ in A1451 and Z0634.



The results of the injections suggest that, even in the worst case, the flux of the radio halos can be underestimated at most of a factor of 1.8 (A1451) and 2.2 (Z0634) at 1.5 GHz, meaning that both the radio halos would be still $4-5$ times below the $P_{1.4GHz}-M_{500}$ correlation. We also note that at these flux levels, the morphology of the injected halos that is recovered from the maps may differ from that of the injected models. For this reason it is impossible to firmly establish weather the halo in A1451 extends only in the Southern region or if there is some emission extending also in the Northern part. Finally, the injection procedure allows us to estimate the size of halos: we found that injected halo models with $r_e\sim140$ kpc and $r_e\sim 130$ kpc (eq.2) match the observed extension of the radio halos in A1451 and Z0634 when the recovered fluxes match the observed ones. 

\subsection{Comparison with the current $P_{1.4GHz}-M_{500}$ correlation for radio halos}
\label{Sec:disc_RH}

Using statistical samples of 30-40 galaxy clusters it has been shown that their radio properties on the Mpc scale have a bimodal distribution: either they host giant radio halos following the correlation between the cluster mass (or X-ray luminosity) and the radio luminosity, or they do not show diffuse radio emission at the sensitivity level of current observations \citep[e.g.][]{cassano13}. This bimodality is connected to the dynamical state of clusters, with radio halo clusters undergoing merging events and non radio halo clusters being generally relaxed. Since its discovery, it was proposed that a bimodal segregation results from the fact that the period of time that clusters spend in the region between the limits and the correlation is smaller than the time they spend on the correlation or in a much fainter state \citep[e.g.][]{brunetti09, donnert13}. In general, this implies that the increase of the statistics would lead to the discovery of an increasing number of underluminous radio halos, sitting in the region between the correlation and the upper limits, which essentially sets the sensitivity of current observations. For illustration, we show in Fig. \ref{Fig:P-M} the position of A1451 and Z0634 compared to the 95\% confidence level region filled by giant radio halos in \citet{cassano13}. In order to detect underluminous radio halos, very deep observations of large samples of galaxy clusters are needed. Our sample of 75 clusters with $M_{500}\gtrsim6\times10^{14}M_\odot$ and $z=0.08-0.33$ (Sect. \ref{Sec:intro}) allows a unique statistical assessment of the distribution of clusters in the radio power-mass diagram and of the occurrence of radio halos in clusters. Here we anticipate that 30 radio halos are found in this sample, 7 of them sit in the region between the current radio-mass correlations and the upper limits (Cuciti et al. in prep.). Although these underluminous halos are statistically subdominant, their presence confirms that the increase of the statistics allows to populate the region below the correlation in the radio-mass diagram, which also leads to an apparent increase of the scatter of such correlation toward lower luminosities. In fact there are reasons to expect a fairly large scatter in the scaling relation between the radio halo power and the cluster masses, due to the superposition of different intermediate radio halo states that are generated by the complex hierarchy of merger events.

One possibility proposed in the literature for underluminous radio halos is that they may have ultra steep spectra, resulting from less energetic merger events (smaller masses or minor mergers in massive systems) or in the case of higher redshift clusters, where inverse Compton losses become severe compared to synchrotron losses \citep{CBS06,brunetti08nature}. Radio halos in the very early or very late stages of their lifetime are also expected to be underluminous and have particularly steep spectra \citep{donnert13}. Regarding A1451 and Z0634, these scenarios are disfavoured by the evidence that they do not host radio halos with very steep spectra (Sect. \ref{Sec:test}).

For the two underluminous radio halos presented in this paper we propose two possibilities. They could be small radio halos, originated in minor mergers where the bulk of turbulence is generated and dissipated in small volumes. The fact that the X-ray morphology of these two clusters is not extremely disturbed may suggest that they are undergoing minor mergers. In this case, the bulk of turbulence may be generated and dissipated in small volumes resulting in relatively flat radio spectra, despite the relatively low energy involved in the mergers events. Alternatively, they could be ``off-state'' halos where the emission is primarily maintained by the continuous injection of secondary electrons of hadronic interactions in the ICM. Possible detections of ``off-state'' radio halos have been reported in the literature \citep[e.g.][]{brown11,bonafede14}. The relatively flat spectra and the small sizes of these two newly discovered radio halos would be in line with both these scenarios.

\subsection{The candidate radio relic}
 
The elongated and arc-like morphology of the source detected North of A1451 (Fig. \ref{Fig:A1451_relic}), together with its spectral index ($\alpha\approx-1.1$, see Sect \ref{Sec:relic}) suggest that it could be a radio relic. With a radio power of $P_{1.4GHz}=1.13\pm0.06\times10^{24}$ W/Hz (Sect. \ref{Sec:relic}), this source would follow the radio-luminosity--mass correlation found by \citet{degasperin14} for radio relics. However, if this source is at the cluster redshift, its projected distance from the cluster center is $\sim3$ Mpc, which would make it the most distant relic from the cluster center ever detected \citep{vazza12,degasperin14}. The virial radius of A1451 is $R_v\sim2.5$ Mpc \footnote{To derive the virial radius, $R_v$, we converted $M_{500}$
to $M_{vir}$, by assuming an NFW profile (e.g. Navarro et al. 1997)
for the dark matter halos and the concentration-mass relation
in Duffy et al. (2008).}, moreover weak lensing studies show that its mass distribution is elongated in the N-S direction \citep[see Fig. 3 in][]{cypriano04}, thus supporting the idea that we would be observing a radio relic located in the cluster very-external regions.
The candidate radio relic in A1451 might be associated to an external shock, rather than a merger shock. Accretion/external shocks are the result of the continuous accretion of matter at several Mpc of distance from the cluster center. They are typically much stronger than merger shocks, since they develop in cold regions where the medium is much cooler than in clusters. On the other hand, they propagate through regions with extremely low gas densities. The energy flux dissipated at these shocks (both into heating of the ICM and CR acceleration) is much smaller than that of merger shocks that are generally faster and propagate into denser ICM \citep{ryu03,vazza09}. This makes external shocks very difficult to detect, unless they encounter and re-accelerate some clouds of relativistic plasma during their propagation. For instance, the candidate relic in A1451 could be originated by an external shock that propagates through the lobes of a dead radio galaxy \citep{ensslin01}. Source S (Fig. \ref{Fig:A1451_relic}) might be responsible for the presence of such ghost plasma. Unfortunately no spectroscopic redshift is available for this source. The 2 MASS catalogue \citep{2mass} reports a $K_s$ magnitude of 15.03. Using the $K_s-z$ relation by \citet{willott03} we estimated that the redshift of source S is $z=0.3\pm0.1$, which does not exclude the possibility that it is close to the cluster redshift. We note that this scenario is in agreement with the fact that the candidate relic is well below the LLS--cluster center distance correlation by \citet{degasperin14}. Such correlation has been found for double radio relics, associated to merger driven shocks. In case of shocks crossing a pre-existing cloud of relativistic plasma, the size of the relic is determined by the size of that cloud which, in general, is expected to extend much less than the surface crossed by external shocks. Clearly, another possibility is that this source is a background giant radio galaxy, although the lack of a significant spectral gradient along the E-W direction disfavours this hypothesis. 

Future studies on the polarization properties of this object will be crucial to establish its nature.

\subsection{Head tail}
\label{Sect:HT}
   \begin{figure}
   \centering
   \includegraphics[scale=0.4]{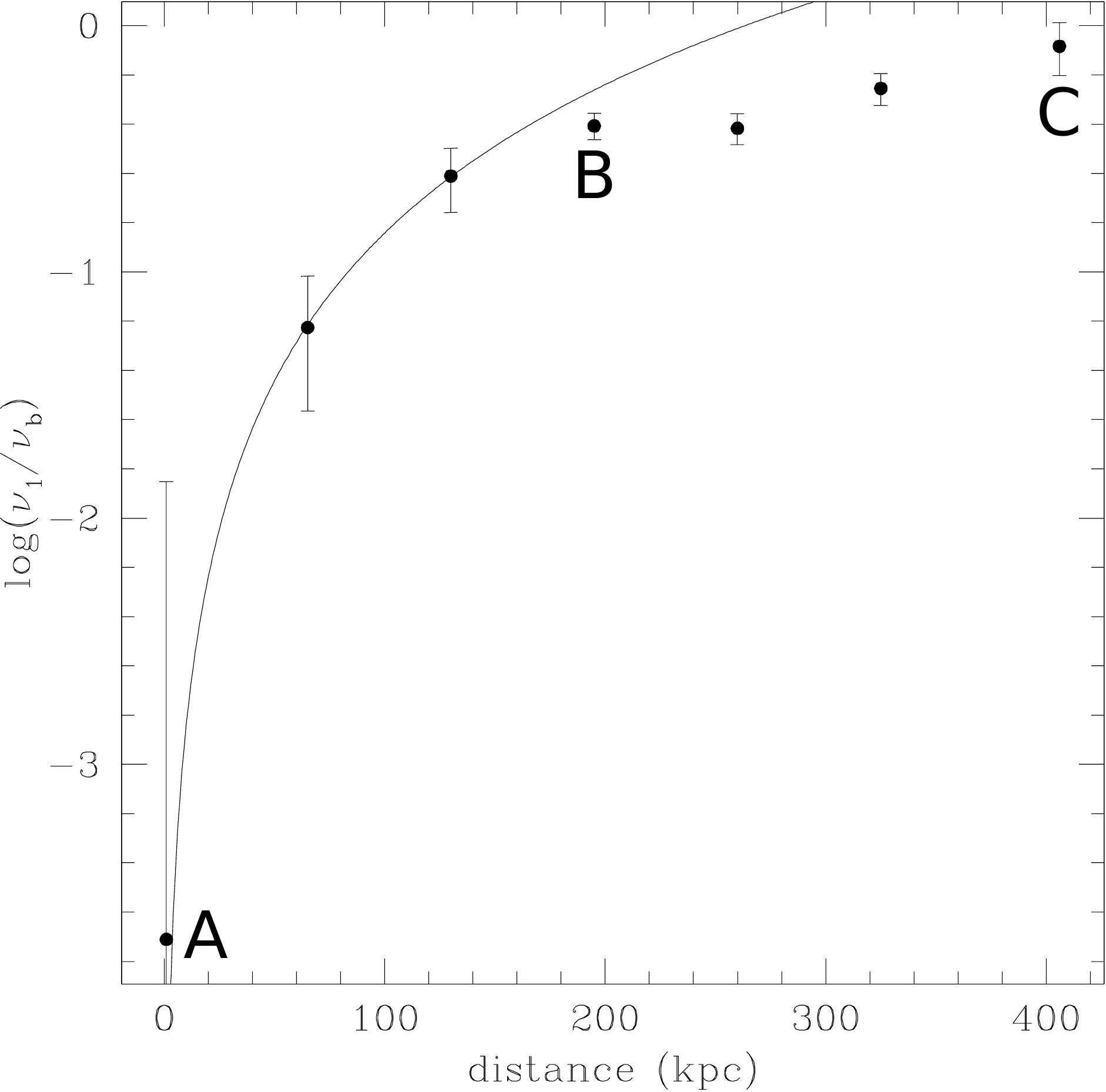}
      \caption{$\nu_{320MHz}/\nu_b$ vs. distance from the ``head'' of the HT galaxy. The line represents $\nu_b\propto d^{-2}$. For display purposes the curve is normalised to the third data point.
      }
\label{Fig:ageing}
   \end{figure}
   
The head tail radio galaxy in Z0634 shows very interesting properties. The spectral index of the tail between 320 and 1500 MHz shows a continuous steepening from $\sim-0.65$ to $\sim-2.4$. In order to understand if such a steepening is consistent with pure radiative ageing of the electrons, we attempt to extract both the spectral index and the surface brightness with distance.
We selected six circular regions along the tail and a final rectangular one (Fig.\ref{Fig:coda_profile}). The size of the circular regions is $\sim1.5$ times the beam area. We selected a rectangular larger area on the last portion of the tail, that is only marginally visible at 1.5 GHz, to have a better signal to noise ratio. We did not consider the very last part of the tail (West of C in Fig. \ref{Fig:coda_profile}, left) because it is only visible at 320 MHz with a poor signal to noise ratio.

\begin{figure*}
   \centering
   \includegraphics[scale=0.4]{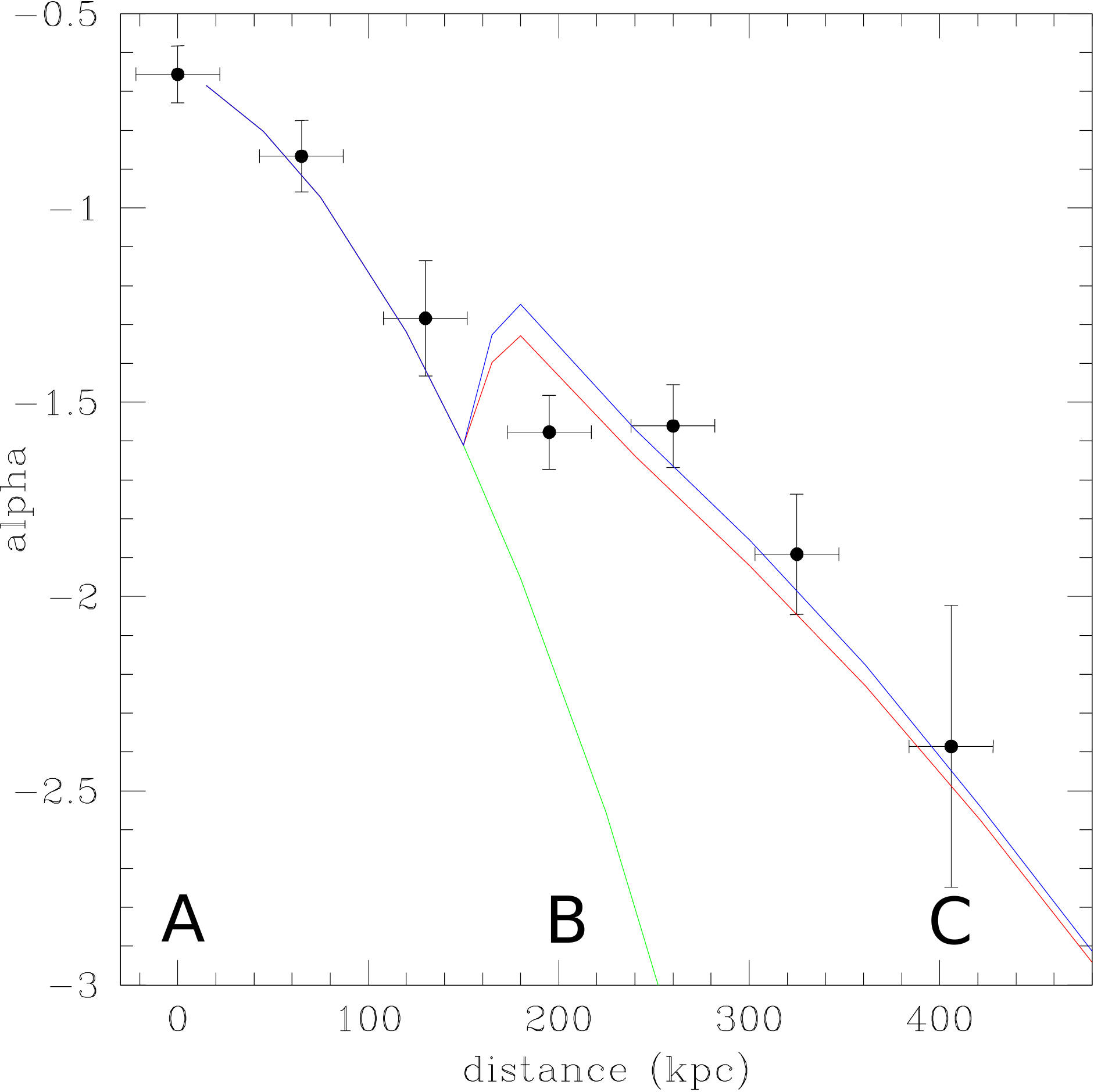} \hspace{1cm}
   \includegraphics[scale=0.4]{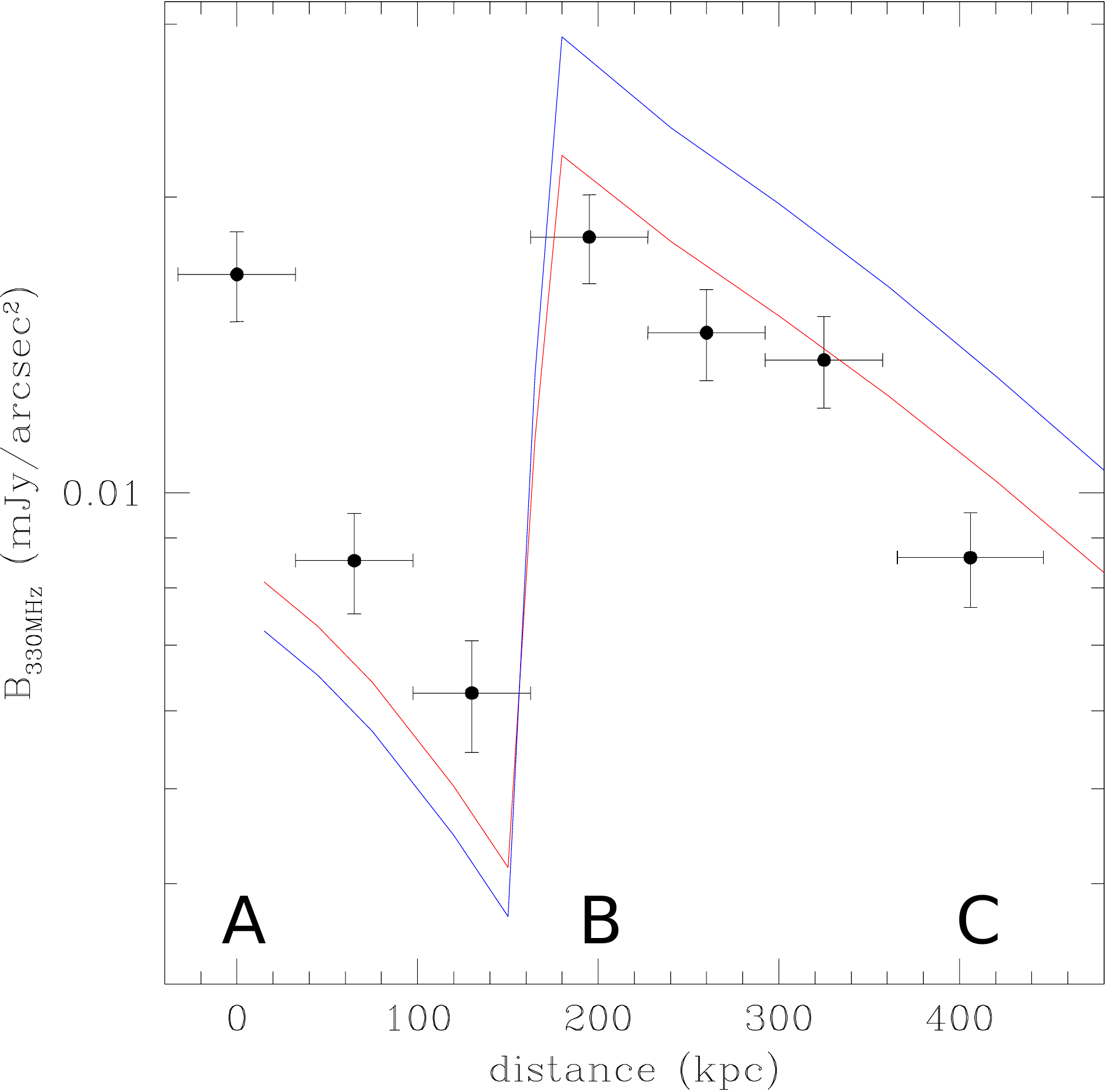}
      \caption{\textit{Left}: spectral index of the HT vs distance from the ``head''. The green line is the classical model of synchrotron ageing of a population of electrons that have been accelerated in the ``head''. The injection spectrum is assumed to match the spectrum in the head of the tail. \textit{Right}: radio surface brightness vs. distance; model normalization ($K_e$ in Appendix) is chosen for best displaying the comparison with the data. A, B and C refer to Fig. \ref{Fig:Z0634_codaspix}. In both panels the model parameters are: $\mathcal M=1.7$, $\theta_{SO}=\pi/2$, $a=1$, $c_S=1330$ km/s, $\sigma_v=1400$ km/s, $\psi(t=0)=1$, $\eta\sin \theta_O=1$, with $\theta_O=60^{\circ}$ and $\theta_{SO}=90^{\circ}$, $\Gamma=5/3$ (red line) and $\Gamma=4/3$ (blue line).
              }
\label{Fig:coda_grafici}
   \end{figure*}

The radio surface brightness and the spectral index show a peculiar evolution along the tail (Fig. \ref{Fig:coda_grafici}). In the first part of the tail (A$\rightarrow$B) they progressively decline as expected due to radiative losses of a population of relativistic electrons that have been accelerated in the head of the radio galaxy (in A). In the presence of synchrotron and IC losses only, the (emitted) break frequency is:
\begin{equation}\label{eq:nub}
\begin{split}
\nu_b\mathrm{(GHz)}=2.6\dfrac{\tau\mathrm{(Gyr)}^{-2}B\mathrm{(\mu G)}}{(B\mathrm{(\mu G)}^2+B_{CMB}\mathrm{(\mu G)}^2)^2}\propto\\
\propto\dfrac{(\eta\sin\theta_O\sigma_v)^2}{B_{CMB}^3}\dfrac{\psi}{(\frac{\psi^2}{3}+1)^2}d_\perp^{-2}
\end{split}
\end{equation}
where $\tau$ is the radiative age of the electrons, $B_{CMB}$ is the equivalent magnetic field of the CMB, $\sigma_v$ is the velocity dispersion of the cluster ($\sigma_v \sim 1400$ km/s) and $\eta\sin\theta_O \sigma_v$ is the (projected) velocity of the head. We note that $\eta\sin\theta_O$ combines two unknowns: the velocity of the HT and the inclination of its orbit. In Eq. \ref{eq:nub}, we have introduced the quantity $B=\psi B_{cmb}/\sqrt{3}$.  $\psi$ parametrises the magnetic field of the tail. For $\psi=1$ the electron emitting at the observing frequencies in a magnetic field $B$ are in the situation of minimum radiative losses and consequently maximum lifetime.
In the simplest interpretation, the fact that the break frequency scales with $d^{-2}$ between A and B (Fig. \ref{Fig:ageing}) is suggestive that adiabatic losses and strong variations of $B$ do not play an important role at this stage, otherwise a faster increase of $\nu_1/\nu_b$ should be observed. More quantitatively, assuming $\psi=1$ and $\eta \sin\theta_0=1$, in Fig. \ref{Fig:coda_grafici} we show the measured spectral index and brightness as a function of projected distance. Also in this case, a simple scenario of pure radiative losses in the region A--B can explain the data. We note that there is degeneracy in the parameters: from Eq. \ref{eq:nub}, one finds that the same match can be obtained with different choices of the parameters $\eta$, $\psi$, i.e. $B$, and $\theta_O$ under the condition $\eta^2 \sin^2\theta_O \frac{\psi}{(\psi^2/3+1)^2}=9/16$.


In B (Fig. \ref{Fig:coda_profile}) the surface brightness is boosted up by a factor of 3.5-4, while the spectral index stops declining and remains almost constant for $\sim 100$ kpc, corresponding to $\sim60$ Myr (if we consider $\eta\sin \theta_O=1$) \footnote{We point out that these values are just indicative, since they depend on the sampling adopted in Fig. \ref{Fig:coda_profile}}. Then, from B to the end of the tail (C), both the surface brightness and the spectral index of the tail restart decreasing and steepening, respectively.
The possibility that we are seeing two HT chasing each other is excluded by the B array image (Fig. \ref{Fig:Z0634_Chandra_D}, right panel), where the bright spot is resolved and appears as part of the diffuse emission of the tail. Moreover, the spectral index in B is very steep and inconsistent with that of synchrotron radiation from a young plasma. On the other hand, the properties of the tail suggest that around the region B the electrons in the tail are currently experiencing re-energization and possibly that the magnetic field in the tail is also amplified.

\begin{figure*}
   \centering
   \includegraphics[width=\hsize]{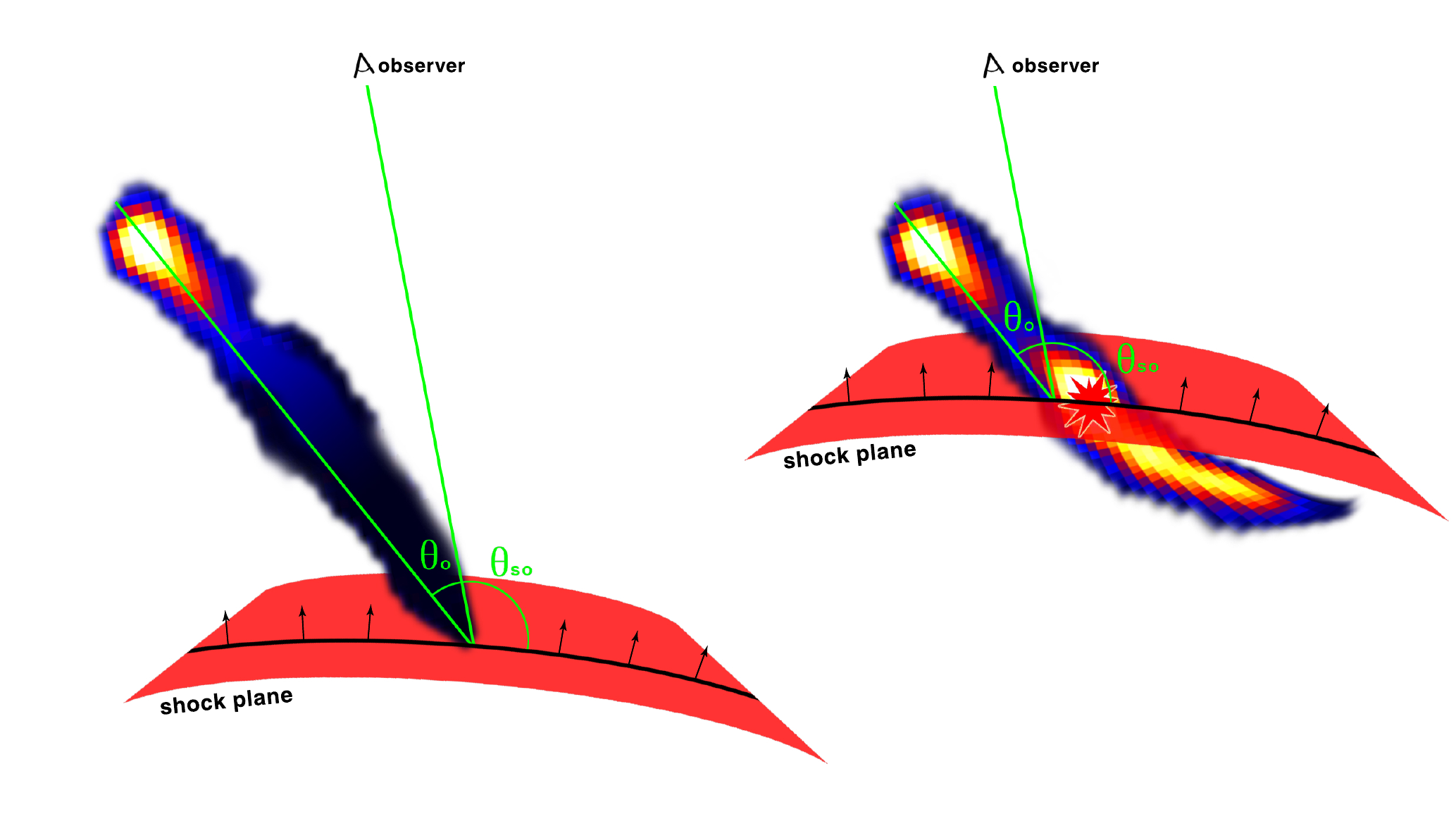}
      \caption{Cartoon of the interaction between the shock front and the HT. Before the interaction (\textit{left}) the surface brightness of the HT progressively declines from the head to the tail. The right panel represents the current situation: the shock has passed through the last part of the tail that appears brighter because of the re-energization that electrons have undergone. The bright spot is where the tail is currently shocked. The shock passage also caused the bending of the tail in the direction parallel to the shock itself.   
              }
\label{Fig:coda_cartoon}
   \end{figure*}

The most natural mechanism that may explain these properties in the ICM is the interaction between a shock and the radio tail \citep[e.g.][]{pfrommer-jones11}. A schematic view is illustrated in Fig. \ref{Fig:coda_cartoon}. The HT galaxy is travelling outward from the cluster center with an angle $\theta_O$ to the line of sight. A shock front impacts the tail first and gets to the point B where we are currently observing it. When the shock passes through the tail the adiabatic compression re-energises a population of aged electrons and increases the magnetic field. This re-energization leads to an increase of the synchrotron brightness in B and affects the observed spectral properties. In the downstream region (B$\rightarrow$C) the electrons have restarted to cool down, in the new magnetic field, but the ageing time has to be computed since the shock passage and it does not depend on the velocity of the tail any longer. Of course, the dynamics of the HT--shock interaction can be very complicated and would require ad hoc simulations that are beyond the aims of this work. For example, the shock crossing may change the inclination of the shocked portion of the tail, implying that projection effect may play an important role. In addition, after the shock passage, the tail may experience some expansion together with the gas downstream, and adiabatic expansion may induce a strong spectral steepening in the tail at later stages due to adiabatic losses and the decline of the magnetic field.

The system is complicated and unfortunately our observations cannot constrain all model parameters. Here we limit ourselves to demonstrate that, in the case of a weak shock travelling at large angle with respect to the plane of the sky, this scenario has the potential to explain the observed properties of the tail (see Appendix for the formalism). To do that, we assume the simplest case where the shock is moving along the line-of-sight and where $\psi(t=0)=1$ (i.e. electrons upstream in the tail have minimum energy losses), $\eta\sin\theta_O=1$ and the velocity dispersion of the cluster $\sigma_v=1400$ km/s. We also assume that particles are continuously injected in the core region of the head tail at a rate that is constant with time (Appendix). Fig. \ref{Fig:coda_grafici} (left panel) shows that this situation matches the observed spectrum upstream (A$\rightarrow$B), with the degeneracy of the parameters upstream given by Eq. \ref{eq:nub}. The evolution of the magnetic filed in the downstream region is given by Eq. \ref{eq:C}, considering $B(t)=B(t_0)C_t^{2/3}$ and the conversion from time to projected distance is given by Eq. \ref{eq:dperp}. The evolution of particles downstream subject to adiabatic compression and (time-variable) radiative losses is calculated in the Appendix. In Fig. \ref{Fig:coda_grafici} we show the model results assuming $\mathcal M=1.7$ and $\theta_O=60^{\circ}$, thus the velocity of the tail is $v=\sigma_v/\sin60^{\circ}\approx1600$ km/s \footnote{Note that for a given magnetic field $\psi$ the normalization of the synchrotron brightness depends on the normalization of the electron spectrum, $K_e$ (Appendix), that is a free parameter}.
The two curves assume $\Gamma=4/3$ (relativistic plasma) and $\Gamma=5/3$ (the case where most of the energy of the plasma in the tail is contributed by thermal gas). Given the number of model parameters (Appendix) and the uncertainties on the dynamics of the interaction between the tail and the shock, a full extrapolation of the model parameters is not meaningful at this stage. However, from Fig. \ref{Fig:coda_grafici}, we can conclude that the proposed scenario can naturally explain the observations for reasonable choices of model parameters (i.e. for a velocity of the tail that is of the order of the cluster velocity dispersion), assuming a weak shock with Mach number $< 2$, as most shocks detected in galaxy clusters. In this respect, a simple estimate of the maximum Mach number can be obtained neglecting losses during the shock compression phase \citep[e.g.][see also Appendix]{markevitch05}. For $\Gamma=5/3$ (for $\Gamma=4/3$ the limit is more stringent) and for a shock travelling at large angle to the plane of the sky this is:

\begin{equation}
\mathcal M\leq \left(\dfrac{3}{4\bigg(\frac{I_u}{I_d}\bigg)^{\frac{3}{2\delta}}-1}\right)^{\frac{1}{2}}
\end{equation}

which gives $\mathcal M \leq 1.58$ by assuming the observed brightness jump $I_d/I_u \sim 4$ and a spectrum of the emitting electrons $\delta = 3.5$; the allowed Mach number is slightly larger when radiative losses during compression are properly included (see Fig. \ref{Fig:coda_grafici}) \footnote{We point out that given the observed spectrum and the Mach numbers used in our calculations, even in the case $\Gamma=5/3$ (in this case particles inside the HT will be shocked) the process of shock acceleration is subdominant with respect to compression.}. 

We also note that no shock is clearly visible at the position B in the currently available X-ray image, suggesting that if a shock is present it should be weak and/or moving in a direction that is significantly inclined with respect to the plane of the sky.

\section{Conclusions}
In this paper we have presented deep GMRT 320 MHz observations and JVLA 1.5 GHz observations in B and D array of the galaxy clusters A1451 and Z0634. We summarize our results below:
\begin{itemize}
\item[$\bullet$] We discovered a radio halo at the center of A1451. The radio halo has a LLS of $\sim 750$ kpc ($r_e\sim140$ kpc based on the injection of fake radio halos, Sect. \ref{Sec:test}) and a $k$-corrected radio power at 1.4 GHz $P_{1.4GHz}=6.4\pm0.7\times10^{23}$ W/Hz. We estimated that the spectral index of this radio halo between 320 and 1500 MHz is in the range $-1.3<\alpha<-1.1$.  \\

\item[$\bullet$] We discovered that Z0634 hosts a radio halo. It has a LLS of $\sim 600$ kpc ($r_e\sim130$ kpc based on the injection of fake radio halos, Sect. \ref{Sec:test}) and a $k$-corrected radio power at 1.4 GHz $P_{1.4GHz}=3.1\pm0.2\times10^{23}$ W/Hz. Its spectral index between 320 and 1500 MHz is $-1.3<\alpha<-0.95$.\\

\item[$\bullet$] Both the radio halos in A1451 and Z0634 are underluminous with respect to the radio power--mass correlation found by \citet{cassano13} and \citet{martinezaviles16} (Fig. \ref{Fig:P-M}), similarly to other few clusters reported in the literature \citep{venturi07,venturi13,sommer17,bonafede15}. That region of the $P_{1.4GHz}-M_{500}$ diagram is usually occupied either by upper limits or by ultra steep spectrum radio halos, however, A1451 and Z0634 do not seem to host radio halos with steep spectra (Sect. \ref{Sec:test}). In the framework of current theoretical models, they could be small radio halos generated during minor mergers where turbulence has been dissipated in smaller volumes, or ``off-state'' radio halos originating from hadronic collisions in the ICM.\\

\item[$\bullet$] We detected an elongated diffuse source $\sim15$ arcmin North of the center of A1451 that we classified as a candidate radio relic. This source has a LAS of $\sim6.5$ arcmin and a $k$-corrected radio power at 1.4 GHz $P_{1.4GHz}=1.13\pm0.06\times10^{24}$ W/Hz. Its spectral index between 320 and 1500 MHz is $\alpha=-1.1\pm0.1$. If this source is at the cluster redshift, its projected distance from the cluster center is $\sim3$ Mpc, which would be the largest distance ever measured for a relic \citep[e.g.][]{vazza12,degasperin14}. We propose that this relic could be the result of an accretion/external shock propagating through some clouds of ghost plasma. The fact that this source is well below the LLS-cluster center distance correlation by \citet{degasperin14} is in agreement with the accretion shock scenario. 
Polarization information will be fundamental to address the nature of this object. \\

\item[$\bullet$] Z0634 hosts a head tail radio galaxy with intriguing characteristics. The spectral index between 320 and 1500 MHz steepens from the head to the tail (from $\alpha=-0.65$ to $\alpha=-2.4$), however this steepening is not consistent with a pure radiative ageing of electrons. Moreover, there is a spot of high surface brightness along the tail and the second part of the tail, after the spot, is brighter than the first part, before the spot. We have shown that the interaction of the head tail with a weak shock ($\mathcal M\lesssim2$) moving toward the outskirts of the cluster can explain the observed properties.  

\end{itemize}

\begin{acknowledgements}
The authors thank the referee for useful comments that have improved the presentation of the paper. VC thanks G. Conte who produced the cartoon in Fig. \ref{Fig:coda_cartoon}. VC, GB, RC, TV acknowledge partial support from grant PRIN INAF 2014. RK acknowledges support from the DST-INSPIRE Faculty award by the Department of Science and Technology, Government of India. The National Radio Astronomy Observatory is a facility of the National Science Foundation operated under cooperative agreement by Associated Universities, Inc. We thank the staff of the GMRT that made these observations possible. GMRT is run by the National Centre for Radio Astrophysics of the Tata Institute of Fundamental Research. The scientific results reported in this article are based in part on data obtained from the \textit{Chandra} Data Archive. This research has made use of the NASA/IPAC Extragalactic Database (NED) which is operated by the Jet Propulsion Laboratory, California Institute of Technology, under contract with the National Aeronautics and Space Administration.
\end{acknowledgements}

%
%

\bibliographystyle{aa} 
\bibliography{biblio_virgi} 

\begin{appendix}
\section{}

The evolution of the electron spectrum in the tail assuming radiative losses and compression (expansion) is \citep{kaiser97,ensslin01}:
\begin{equation}
\label{eq:dN/dt}
\dfrac{\partial N}{\partial t}=\dfrac{\partial}{\partial\gamma}\bigg\{ N\bigg[\dfrac{4\sigma_T}{3m_ec}\bigg(\dfrac{B^2}{8\pi}+\dfrac{B_{CMB}^2}{8\pi}\bigg)\gamma^2+ \dfrac{1}{3}\dfrac{\dot{V}}{V}\gamma\bigg]\bigg\}
\end{equation}
where the first term accounts for radiative (synchrotron and IC) losses and the second term for adiabatic compression or expansion.

If electrons have an initial energy distribution in the form $N(\gamma,0)=k_e\gamma^{-\delta}$ (with $\gamma$ between $\gamma_{min}$
and $\gamma_{max}$ such that $\gamma_{min} << \gamma_r << \gamma_{max}$, $\gamma_r$ being the typical energy of the electrons emitting at 330 and 1400 MHz) and if we define the compression factor with time $C_t=V_0/V(t)$, the magnetic field will evolve with time according to $B(t)=B(0)C_t^{2a/3}$, where $a=1$ in the case of isotropic compression$/$expansion and the solution of Eq. \ref{eq:dN/dt} is : 
\begin{equation}
N(\gamma,t)=K_eC_t^{\frac{\delta+2}{3}}\gamma^{-\delta}\bigg(1-\frac{\gamma}{\gamma_{*t}}\bigg)^{\delta-2}
\end{equation}
where 
\begin{equation}
\gamma_{*t}^{-1}=\dfrac{\sigma_tB_{CMB}^2}{6\pi m_ec}\int_0^tdx\bigg(1+\frac{\psi_x^2}{3}\bigg)\bigg(\dfrac{C_x}{C_t}\bigg)^{\frac{1}{3}}
\end{equation}
where we have introduced the quantity
\begin{equation}
\psi_t=\dfrac{B_0}{B_{CMB}/\sqrt{3}}C_t^{\frac{2a}{3}}
\end{equation} 
$\psi_t=1$ is the case where radiative losses of electrons emitting at a given frequency in a given magnetic field are minimized.

Following \citet{ensslin01}, we assume 3 phases: (i) an initial phase where electrons simply age ($t \leq t_1$), (ii) a phase where the shock compresses the tail ($t_1 \leq t \leq t_2$, where $t_2-t_1= \Phi /(\mathcal M c_s)$, $\Phi$ is the thickness of the tail, $c_s$ is the sound speed and $\mathcal M$ is the Mach number), and (iii) a new phase of cooling of the electrons after shock passage ($t > t_2$). Under this assumptions the compression factor is
\begin{equation}\label{eq:C}
  \left\{
  \begin{array}{ll}
   C_t=1 & t\leq t_1\\
   C_t=1+(C_{t_2}-1)\dfrac{t-t_1}{t_2-t_1}& t_1<t\leq t_2\\
   C_t=C_{t_2}=\dfrac{(\Gamma+1)\mathcal M^2}{(\Gamma-1)\mathcal M^2+2}& t>t_2
  \end{array}
  \right.
\end{equation}
$\Gamma$ is the adiabatic index of the gas into the tail. The resulting synchrotron emissivity is:
\begin{equation}
J(\nu,t)=\dfrac{e^3\sqrt{3}}{m_ec^2}\int_0^{\frac{\pi}{2}}d\theta(\sin^2\theta) B\int d\gamma N(\gamma,t)x\int_x^\infty K_{5/3}(z)dz
\end{equation}
where $x=\dfrac{4\pi m_ec\nu}{3eB\sin\theta \gamma^2}$ and $K_{5/3}(z)$ is the 5/3 modified Bessel function.

Given the geometry in Fig. \ref{Fig:coda_cartoon}, the conversion from time to projected distance is given by
\begin{equation}\label{eq:dperp}
  \left\{
  \begin{array}{ll}
   d_\perp=\eta\sigma_v\sin\theta_Ot& t\leq t_2\\
   d_\perp=c_S\mathcal M\dfrac{\sin\theta_O}{\cos\bigg({\theta_{SO}+\theta_O-\frac{\pi}{2}}\bigg)}(t-t_2)& t>t_2
  \end{array}
  \right.
\end{equation}
where $\eta \sigma_V$ is the velocity of the head of the tail. The synchrotron brightness is:
\begin{equation}\label{eq:brightness}
I(\nu,d_{\perp})=\dfrac{1}{2\pi}\int_{d_{\perp}}\dfrac{J(\nu,r(t))r dr}{\sqrt{r^2-d_{\perp}^2}}
\end{equation}
where the integral is done along the tail intercepted by the line of sight.

The model parameters are: the velocity of the HT ($\eta$), the magnetic field ($\psi$), the normalization of the electron spectrum ($K_e$), the geometry of the problem ($\theta_O$, $\theta_{SO}$) and the shock Mach number ($\mathcal M$). In comparing the model with the data, we focus on key observables: the spatial behaviour of the synchrotron spectral index and of the synchrotron brightness. A comparison with these properties is given in Fig. \ref{Fig:coda_grafici} for reference model parameters. In particular, a key observable that constrains the shock Mach number in the model is the observed jump of the synchrotron brightness across the shock, i.e. by the ratio of the downstream and upstream brightness, $I_d/I_u$; $I_d$ and $I_u$ are given by Eq. \ref{eq:brightness} evaluated at time = $t_2$ and $t_1$, respectively.
An useful upper limit to the expected brightness jump can be easily derived by assuming that energy losses during the compression phase can be neglected. In this case, if we assume a power-law spectrum of electrons, $N(\gamma) \propto \gamma^{-\delta}$, the ratio $I_d/I_u = C_{t_2}^{2\delta/3 +1} l_d/l_u$ \citep[e.g.][]{markevitch05} where $l_d$ and $l_u$ are the downstream and upstream thickness intercepted by the line of sight respectively. For reference this gives $I_d/I_u= C_{t_2}^{2\delta/3 +1}$ and $=C_{t_2}^{2\delta/3}$ in the situation of shock travelling at small and large angles to the plane of the sky, respectively. 
\end{appendix}

\end{document}